\pdfoutput=1
\documentclass[cernpreprint, atlasdraft=false, coverpage=false, UKenglish, texmf, orcidlogo]{atlasdoc}
\usepackage{atlaspackage}
\usepackage{atlasbiblatex}

\usepackage{atlasphysics}
\graphicspath{{logos/}{figures/}}
\usepackage{multirow}
\usepackage{ANA-HDBS-2019-08-PAPER-defs}
\usepackage{atlas_tables-defs}

\AtlasTitle{Search for heavy resonances in final states with four leptons and missing transverse momentum or jets in $pp$ collisions at $\sqrt{s} = 13$ TeV with the ATLAS detector}

\AtlasVersion{2.0}

\AtlasAbstract{%
A search for a new heavy boson produced via gluon-fusion in the four-lepton channel with missing transverse momentum or jets is performed. The search uses proton--proton collision data equivalent to an integrated luminosity of 139 fb$^{-1}$ at a centre-of-mass energy of 13 TeV collected by the ATLAS detector between 2015 and 2018 at the Large Hadron Collider. This study explores the decays of heavy bosons: $R\rightarrow SH$ and $A\rightarrow ZH$, where $R$ is a CP-even boson, $A$ is a CP-odd boson, $H$ is a CP-even boson, and $S$ is considered to decay into invisible particles that are candidates for dark matter. In these processes, $S\rightarrow \textrm{invisible}$ and $H\rightarrow ZZ$. The $Z$ boson associated with the heavy scalar boson $H$ decays into all decay channels of the $Z$ boson. The mass range under consideration is 390--1300 (320--1300) GeV for the $R$ ($A$) boson and 220--1000 GeV for the $H$ boson. No significant deviation from the Standard Model backgrounds is observed. The results are interpreted as upper limits at a 95\% confidence level on the cross-section times the branching ratio of the heavy resonances.
}

\author{The ATLAS Collaboration}

\AtlasRefCode{HDBS-2019-08}

\PreprintIdNumber{CERN-EP-2023-291}

\AtlasJournalRef{JHEP 10 (2024) 130}
\AtlasDOI{10.1007/JHEP10(2024)130}

\AtlasTitle{Search for heavy resonances in final states with four leptons and missing transverse momentum or jets in $pp$ collisions at $\sqrt{s} = 13$ TeV with the ATLAS detector}

\AtlasVersion{2.0}

\AtlasAbstract{%
A search for a new heavy boson produced via gluon-fusion in the four-lepton channel with missing transverse momentum or jets is performed. The search uses proton--proton collision data equivalent to an integrated luminosity of 139 fb$^{-1}$ at a centre-of-mass energy of 13 TeV collected by the ATLAS detector between 2015 and 2018 at the Large Hadron Collider. This study explores the decays of heavy bosons: $R\rightarrow SH$ and $A\rightarrow ZH$, where $R$ is a CP-even boson, $A$ is a CP-odd boson, $H$ is a CP-even boson, and $S$ is considered to decay into invisible particles that are candidates for dark matter. In these processes, $S\rightarrow \textrm{invisible}$ and $H\rightarrow ZZ$. The $Z$ boson associated with the heavy scalar boson $H$ decays into all decay channels of the $Z$ boson. The mass range under consideration is 390--1300 (320--1300) GeV for the $R$ ($A$) boson and 220--1000 GeV for the $H$ boson. No significant deviation from the Standard Model backgrounds is observed. The results are interpreted as upper limits at a 95\% confidence level on the cross-section times the branching ratio of the heavy resonances.
}

\author{The ATLAS Collaboration}

\AtlasRefCode{HDBS-2019-08}

\PreprintIdNumber{CERN-EP-2023-291}

\AtlasJournalRef{JHEP 10 (2024) 130}
\AtlasDOI{10.1007/JHEP10(2024)130}

\hypersetup{pdftitle={ATLAS document},pdfauthor={The ATLAS Collaboration}}

\begin{document}

\maketitle

\tableofcontents

\section{Introduction}
\label{sec:intro}

In 2012, the ATLAS and CMS collaborations independently discovered a new particle~\cite{HIGG-2012-27, CMS-HIG-12-028} with properties that are consistent with the Higgs boson proposed by the Standard Model (SM)~\cite{HIGG-2014-06, HIGG-2013-17, CMS-HIG-14-009, CMS-HIG-14-018}. %
It is still debated whether the Higgs boson is a standalone particle or part of an extended Higgs sector suggested by, e.g., the two-Higgs-doublet model (2HDM)~\cite{Branco:2011iw, Lee:1973iz}. The 2HDM predicts the existence of five Higgs bosons, including a CP-even particle similar to the SM Higgs boson ($h$), a heavier CP-even Higgs boson ($H$), a CP-odd particle ($A$), and charged Higgs bosons ($H^{\pm}$). Another extension of the SM is the 2HDM+S model~\cite{vonBuddenbrock:2016rmr, Muhlleitner:2016mzt}, which introduces a scalar boson ($S$) in addition to the 2HDM particles. The $S$ boson is assumed to be a dark matter portal and a potential source of missing transverse momentum, achieved through its decay to a pair of dark matter candidates ($\chi\overline{\chi}$)~\cite{vonBuddenbrock:2016rmr}.

\begin{figure*}
\begin{center}
\subfloat[]{\includegraphics[width=0.31\linewidth,height=0.14\textheight]{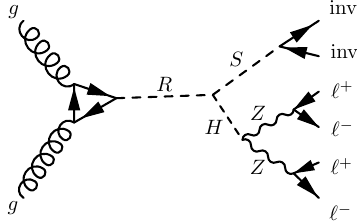}\label{fig:RSH}}\hfill
\subfloat[]{\includegraphics[width=0.31\linewidth,height=0.14\textheight]{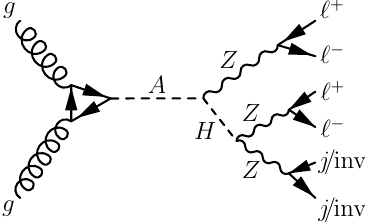}\label{fig:H2l}}\hfill
\subfloat[]{\includegraphics[width=0.31\linewidth,height=0.14\textheight]{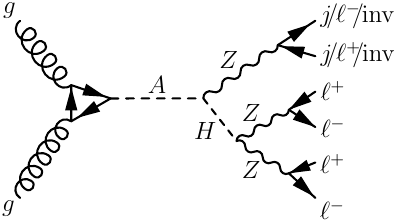}\label{fig:H4l}}
\end{center}
\vspace{-0.5cm}
\caption{Feynman diagrams representing the production of heavy bosons via gluon-fusion at leading-order for the \protect\subref{fig:RSH}  $R\rightarrow SH\rightarrow 4\ell+E^{\textrm{miss}}_{\textrm{T}}$ signal, \protect\subref{fig:H2l} $A\rightarrow Z(\rightarrow 2\ell)H(\rightarrow 2\ell + jj/\textrm{invisible})$ signal, and \protect\subref{fig:H4l} $A\rightarrow Z(\rightarrow jj/\ell^+\ell^-/\textrm{invisible})H(\rightarrow 4\ell)$ signal, where $\ell$ could either be an electron or a muon, $j$ represents a jet and inv denotes an invisible particle.}\label{fig:Feynman}
\end{figure*}

This paper presents searches for bosonic resonances that decay to four charged leptons ($4\ell$, where $\ell$ is either an electron or a muon), along with missing transverse momentum ($E^{\textrm{miss}}_{\textrm{T}}$) or two jets ($jj$). The study focuses on the mass range of heavy bosons, where the invariant mass of the four leptons ($m_{4\ell}$) is above 200~GeV. This study uses proton--proton collision data at a centre-of-mass energy of 13~TeV and with an integrated luminosity of 139~fb$^{-1}$ collected by the ATLAS detector in 2015--2018 at the Large Hadron Collider (LHC). Two different scenarios are considered for the signal model. First, the 2HDM+S model is extended to cover more general situations for various missing energy magnitudes by adding a heavy scalar $R$, where $R$ decays to $H$ and $S$ bosons with $m_R> m_H+m_S$. The $S$  boson decays into dark matter particles ($S\rightarrow \textrm{invisible}$), and the $H$ decays to $4\ell$ through the decay of two $Z$ bosons ($H\rightarrow ZZ\rightarrow 4\ell$). The phenomenology of the new $R\rightarrow SH$ topology can easily embed into the 2HDM+S model using an approach similar to that of Ref.~\cite{vonBuddenbrock:2016rmr}. The masses of the $R$ and $H$ bosons are varied to control the missing transverse momentum, and the $S$ mass is fixed to 160~GeV, see Section~\ref{sec:data}. Second, a 2HDM-based baryogenesis scenario is considered, which generates matter and antimatter asymmetry. Searches for models consistent with baryogenesis were conducted at the LHC in several channels, such as $H\rightarrow hh$~\cite{ATLAS:2019qdc,CMS:2018ipl,}, $H\rightarrow WW/ZZ$~\cite{ATLAS:2017otj,ATLAS:2017jag,ATLAS:2020fry,ATLAS:2019nat,CMS:2019bnu,CMS:2018amk} and $A\rightarrow Zh$~\cite{ATLAS:2017xel,CMS:2019kca}. Searches in the $A\rightarrow ZH\rightarrow \ell^+\ell^-b\bar{b}/\ell^+\ell^-\tau^+\tau^-$~\cite{CMS:2016xnc,CMS-HIG-18-012,ATLAS:2018oht,ATLAS:2020gxx} and $A\rightarrow ZH\rightarrow \ell^+\ell^-t\bar{t}/\nu\bar{\nu}b\bar{b}$~\cite{HDBS-2021-02} final states were also carried out at the LHC. In these searches, for a strong first-order phase transition to occur in the early universe, the hypothesis $m_A>m_H$ is preferred. Therefore, the $A\rightarrow ZH\rightarrow 4\ell+X$ channel is added to this study to explore regions with four leptons and jet activities. $X$ could represent two leptons, two jets, or a pair of invisible particles, with the assumption that $H$ decays into two Z bosons. In this signal, $A$ is a CP-odd scalar which decays to a CP-even scalar $H$ and a $Z$ boson. Two decay scenarios are considered for the associated production of $Z$ and $H$ bosons: $A\rightarrow Z(\rightarrow \ell^+\ell^-/jj/\textrm{invisible})H(\rightarrow 4\ell)$ and $A\rightarrow Z(\rightarrow 2\ell)H(\rightarrow 2\ell+jj/\textrm{invisible})$, which are combined into one signal referred to as the $A\rightarrow ZH\rightarrow 4\ell+X$ signal. In this analysis, only the gluon--fusion production mode is considered for both the $R\rightarrow SH\rightarrow 4\ell+E^\textrm{miss}_{\textrm{T}}$ and $A\rightarrow ZH\rightarrow 4\ell+X$ signals, as illustrated by the Feynman diagrams of Figure~\ref{fig:Feynman}.

This paper is organised as follows. The ATLAS detector is described in Section~\ref{sec:detector}. Section~\ref{sec:data} presents the data and Monte Carlo samples, followed by the object reconstruction in Section~\ref{sec:objects}. Section~\ref{sec:strategy} describes the analysis strategy, and the signal and background modelling are discussed in Section~\ref{sec:modelling}. Section~\ref{sec:systematic} lists the experimental and theoretical systematic uncertainties. Section~\ref{sec:stat_model} explains the statistical model used in the analysis and the results are discussed in Section~\ref{sec:result}. The conclusion is given in Section~\ref{sec:conclusion}.


%
\section{The ATLAS detector}
\label{sec:detector}

The ATLAS detector~\cite{PERF-2007-01} at the LHC covers nearly the entire solid angle around the collision point.\footnote{\AtlasCoordFootnote} It consists of an inner tracking detector surrounded by a thin superconducting solenoid, electromagnetic and hadron calorimeters, and a muon spectrometer incorporating three large superconducting air-core toroidal magnets.

The inner-detector system (ID) is immersed in a \qty{2}{\tesla} axial magnetic field and provides charged-particle tracking in the range \(|\eta| < 2.5\). The high-granularity silicon pixel detector covers the vertex region and typically provides four measurements per track, the first hit normally being in the insertable B-layer (IBL) installed before Run~2~\cite{PIX-2018-001}. It is followed by the silicon microstrip tracker (SCT), which usually provides eight measurements per track. These silicon detectors are complemented by the transition radiation tracker (TRT), which enables radially extended track reconstruction up to \(|\eta| = 2.0\). The TRT also provides electron identification information based on the fraction of hits (typically 30 in total) above a higher energy-deposit threshold corresponding to transition radiation.

The calorimeter system covers the pseudorapidity range \(|\eta| < 4.9\). Within the region \(|\eta|< 3.2\), electromagnetic calorimetry is provided by barrel and endcap high-granularity lead/liquid-argon (LAr) calorimeters, with an additional thin LAr presampler covering \(|\eta| < 1.8\) to correct for energy loss in material upstream of the calorimeters. Hadron calorimetry is provided by the steel/scintillator-tile calorimeter, segmented into three barrel structures within \(|\eta| < 1.7\), and two copper/LAr hadron endcap calorimeters. The solid angle coverage is completed with forward copper/LAr and tungsten/LAr calorimeter modules optimised for electromagnetic and hadronic energy measurements respectively.

The muon spectrometer (MS) comprises separate trigger and high-precision tracking chambers measuring the deflection of muons in a magnetic field generated by the superconducting air-core toroidal magnets. The field integral of the toroids ranges between \num{2.0} and \qty{6.0}{\tesla\metre} across most of the detector. Three layers of precision chambers, each consisting of layers of monitored drift tubes, cover the region \(|\eta| < 2.7\), complemented by cathode-strip chambers in the forward region, where the background is highest. The muon trigger system covers the range \(|\eta| < 2.4\) with resistive-plate chambers in the barrel, and thin-gap chambers in the endcap regions. Interesting events are selected by the first-level trigger system implemented in custom hardware, followed by selections made by algorithms implemented in software in the high-level trigger~\cite{TRIG-2016-01}. The first-level trigger accepts events from the \qty{40}{\MHz} bunch crossings at a rate below \qty{100}{\kHz}, which the high-level trigger further reduces in order to record events to disk at about \qty{1}{\kHz}. An extensive software suite~\cite{ATL-SOFT-PUB-2021-001} is used in data simulation, in the reconstruction and analysis of real and simulated data, in detector operations, and in the trigger and data acquisition systems of the experiment.


%
\section{Data and simulated event samples}
\label{sec:data}

The data used in this analysis were produced through proton--proton (\(pp\)) collisions at a centre-of-mass energy of 13~TeV recorded by the ATLAS detector at the LHC from 2015 to 2018. Events are required to satisfy data-quality requirements, and after applying the event cleaning criteria, the total integrated luminosity of the entire data set is 139~fb$^{-1}$ \cite{DAPR-2018-01}. The data events in the four-lepton final state were recorded using a combination of single-lepton, dilepton, and trilepton triggers~\cite{TRIG-2018-05,TRIG-2018-01}. The single-lepton triggers for isolated muons featured $p_{\textrm{T}}$ thresholds ranging from 20 to 60~GeV and 50 GeV for muons without isolation requirement. In contrast, dilepton triggers had thresholds from 8 to 22~GeV, and trilepton triggers employed thresholds between 4 and 20~GeV. For electrons, the single-lepton triggers encompassed $p_{\textrm{T}}$ thresholds ranging from 24 to 300~GeV, with dilepton triggers varying from 12 to 24~GeV and trilepton triggers featuring thresholds between 9 and 24~GeV. Additionally, when a dilepton trigger encountered an event with an electron and a muon, the $p_{\textrm{T}}$ thresholds for electrons ranged from 7 to 26~GeV. For muons, the thresholds varied from 8 to 24~GeV. These $p_{\textrm{T}}$ thresholds exhibited variations across the data-taking period from 2015 to 2018 due to rising luminosity~\cite{ATL-DAQ-PUB-2016-001,ATL-DAQ-PUB-2017-001}, but the overall trigger efficiency was about 98\%.

Signal and background events were generated according to ATLAS detector configurations and utilised for signal optimisation, background parametrisation, and systematic uncertainty estimation. Each Monte Carlo (MC) generator produced events that underwent ATLAS detector simulation~\cite{SOFT-2010-01} within the \textsc{Geant4} framework~\cite{GEANT4:2002zbu}. The effect of multiple interactions in the same and neighbouring bunch crossings (\pileup) was modelled by overlaying the simulated hard-scattering event with inelastic \(pp\) events generated with \textsc{Pythia 8.186}~\cite{Sjostrand:2007gs} at leading-order (LO) using the \textsc{NNPDF2.3} LO set of parton distribution functions (PDFs)~\cite{Ball:2012cx} and the A3 set of tuned parameters (tune)~\cite{ATL-PHYS-PUB-2016-017}.

The $q\overline{q}\rightarrow ZZ$ background process was simulated with \textsc{Sherpa 2.2.2}~\cite{Bothmann:2019yzt} using matrix elements at next-to-leading-order (NLO) accuracy in QCD for up to one additional parton and at LO accuracy for up to three additional parton emission. The virtual QCD corrections were provided by the \textsc{OpenLoops} library~\cite{Buccioni:2019sur,Cascioli:2011va,Denner:2016kdg}. The $gg\rightarrow ZZ$ background process was generated by \textsc{Sherpa 2.2.2}, including off-shell effects and Higgs boson contributions, using LO-accurate matrix elements for up to one additional parton emission. The electroweak $q\overline{q}\rightarrow ZZ$ (EW) events, consisting of four leptons and two jets, were simulated using \textsc{Sherpa 2.2.2}. In addition, the $VVV$ background events, including processes such as $ZZZ$, $ZZW$, and $WWZ$ with at least four prompt charged leptons, were simulated using \textsc{Sherpa 2.2.2}. The matrix element calculations for these processes were matched and merged with the \textsc{Sherpa}  parton shower based on Catani--Seymour dipole factorisation~\cite{Gleisberg:2008fv,Schumann:2007mg} using the \textsc{MEPS@NLO} prescription~\cite{Hoeche:2011fd,Hoeche:2012yf,Hoeche:2009rj}. The production of $Z$+jets was simulated with the
\textsc{Sherpa 2.2.1} generator using NLO matrix elements for up to two partons, and LO matrix elements for up to four partons calculated with the Comix~\cite{Gleisberg:2008fv}
and \textsc{OpenLoops} libraries. All backgrounds simulated with \textsc{Sherpa} use \textsc{NNPDF3.0} based on next-to-next-to-leading-order (NNLO) set of PDFs~\cite{Ball:2014uwa}, along with the dedicated set of tuned parton shower parameters developed by the \textsc{Sherpa}  authors.

Events containing charged leptons coming from the $t\bar{t}V$ background process ($V=Z\,\textrm{or}\, W^{\pm}$) were modelled using the \textsc{MadGraph5\_aMC@NLO 2.3.3}~\cite{Alwall:2011uj} generator at NLO with the \textsc{NNPDF3.0 NLO} PDF. The events were interfaced to \textsc{Pythia 8.210} using the A14 tune~\cite{ATL-PHYS-PUB-2014-021} and the \textsc{NNPDF2.3 LO} PDF set. The $t\bar{t}$ background was generated using \textsc{Powheg-Box v2}~\cite{Alioli:2010xd,Nason:2004rx,Frixione:2007vw,Alioli:2010xd} at NLO with the \textsc{NNPDF3.0 NLO} PDF set. The events were interfaced to \textsc{Pythia 8.230} to model the parton shower, hadronisation, and underlying event, with parameters set according to the A14 tune and using the \textsc{NNPDF2.3 LO} set of PDFs. The \textsc{Powheg-Box v2} generator
was used to simulate the \(WZ\)~\cite{Nason:2013ydw} production
process at NLO accuracy in QCD. Events were interfaced to \textsc{Pythia 8.230} for the modelling of the parton shower, hadronisation, and underlying event, with parameters set according to the \AZNLO tune~\cite{STDM-2012-23}. The \CT[10] PDF set~\cite{Lai:2010vv} was used
for the hard-scattering processes, whereas the \CTEQ[6L1] PDF set~\cite{Pumplin:2002vw} was used for the parton shower. %
The decays of the bottom and charm hadrons for the top-quark processes were simulated using version 1.2 of the \textsc{EvtGen} program~\cite{Lange:2001uf}, except for using version 1.6 for $t\bar{t}$ production.

The $R\rightarrow SH\rightarrow 4\ell+E^{\textrm{miss}}_{\textrm{T}}$ signal was simulated using \textsc{Pythia 8.244} with the \textsc{A14} tune and \textsc{NNPDF2.3 LO} PDF set. The $A\rightarrow Z(\rightarrow jj/\ell^+\ell^-/\textrm{invisible})H(\rightarrow 4\ell)$ and $A\rightarrow Z(\rightarrow2\ell)H(\rightarrow 2\ell + jj/\textrm{invisible})$ signals were generated by \textsc{MadGraph5\_aMC@NLO 2.7.2} at LO accuracy with the \textsc{NNPDF2.3} LO PDF set, which is interfaced to \textsc{Pythia 8.244} for parton showering and
hadronisation with the \textsc{A14} tune. The $R$ mass is considered in the range of 390--1300~GeV, with the $S$ mass fixed to 160~GeV, while the $A$ mass is in the range of 320--1300~GeV. The $H$ mass value for all signal processes is in the range of 220--1000~GeV. The assumption of the $S$ mass is motivated by the phenomenology study presented in Ref.~\cite{vonBuddenbrock:2017gvy}. However, the effect of this choice on the distribution of the missing transverse momentum was studied and it was found that the $S$ mass only affects the distribution if its mass is above 200~GeV. Therefore, fixing the $S$ mass reduces the free parameters and simplifies the analysis. Signals were generated using the narrow width approximation (NWA)~\cite{Berdine:2007uv} for both $R$ and $H$ in the $R\rightarrow SH\rightarrow 4\ell+E^{\textrm{miss}}_{\textrm{T}}$ signal, with the width of $R$ and $H$ set to 10 MeV. Similarly, for the $A\rightarrow ZH\rightarrow4\ell+X$ signal, the narrow width approximation was employed with widths of 1~MeV for both $A$ and $H$.
Two mass points, $(m_A, m_H) = (320, 220)$~GeV and $(m_A, m_H) = (1190, 600)$~GeV, were also generated using the large width approximation (LWA) to evaluate the impact of non-negligible natural widths. Table~\ref{samples} summarises the generators, shower model, matrix element accuracy, tune, and PDF sets used in signal and background simulation.
\begin{table}[tbp]
\begin{center}
\resizebox{\textwidth}{!}{%
\begin{tabular}{lccc}
\toprule
Process & Generator and Parton shower & QCD accuracy  & Tune and PDF       \\
\midrule
$R\rightarrow SH\rightarrow 4\ell+E^{\textrm{miss}}_{\textrm{T}}$  & \textsc{Pythia 8.244}~\cite{Sjostrand:2014zea} & LO & A14 and \textsc{NNPDF2.3 LO}~\cite{Ball:2011mu,ATL-PHYS-PUB-2014-021}\\
$A\rightarrow Z(\rightarrow jj/\ell^+\ell^-/\textrm{invisible})H(\rightarrow 4\ell)$  & \textsc{MadGraph5\_aMC@NLO 2.7.2} + \textsc{Pythia 8.244}~\cite{Alwall:2011uj,Sjostrand:2014zea} & LO & A14 and \textsc{NNPDF2.3 LO}~\cite{Ball:2011mu,ATL-PHYS-PUB-2014-021}\\
$A\rightarrow Z(\rightarrow 2\ell)H(\rightarrow 2\ell+jj/\textrm{invisible})$  & \textsc{MadGraph5\_aMC@NLO 2.7.2} + \textsc{Pythia 8.244}~\cite{Alwall:2011uj,Sjostrand:2014zea} & LO & A14 and \textsc{NNPDF2.3 LO}~\cite{Ball:2011mu,ATL-PHYS-PUB-2014-021}\\
$t\overline{t}V$ ($V = W/Z$) & \textsc{MadGraph5\_aMC@NLO 2.3.3} + \textsc{Pythia 8.210}~\cite{Alwall:2011uj,Sjostrand:2014zea} & NLO+LO & A14 and \textsc{NNPDF3.0 NLO} \& \textsc{NNPDF2.3 LO}~\cite{Ball:2014uwa,Ball:2011mu,ATL-PHYS-PUB-2014-021}\\
$q\overline{q}\rightarrow ZZ$  & \textsc{Sherpa 2.2.2} + \textsc{MEPS@NLO}~\cite{Hoeche:2011fd,Hoeche:2012yf,Hoeche:2009rj} & NLO (0- and 1-jet), LO (2- and 3-jet) & \textsc{Sherpa} and \textsc{NNPDF3.0 NNLO}~\cite{Ball:2014uwa}\\
$gg\rightarrow ZZ$                & \textsc{Sherpa 2.2.2} + \textsc{MEPS@NLO}~\cite{Hoeche:2011fd,Hoeche:2012yf,Hoeche:2009rj} & LO (0- and 1-jet)      & \textsc{Sherpa} and \textsc{NNPDF3.0 NNLO}~\cite{Ball:2014uwa}\\
$q\overline{q}\rightarrow ZZ$ (EW) & \textsc{Sherpa 2.2.2} + \textsc{MEPS@NLO}~\cite{Hoeche:2011fd,Hoeche:2012yf,Hoeche:2009rj} & LO & \textsc{Sherpa} and \textsc{NNPDF3.0 NNLO}~\cite{Ball:2014uwa} \\
$ZZZ (4\ell2\nu,6\ell0\nu)$ & \textsc{Sherpa 2.2.2} + \textsc{MEPS@NLO}~\cite{Hoeche:2011fd,Hoeche:2012yf,Hoeche:2009rj} & NLO & \textsc{Sherpa} and \textsc{NNPDF3.0 NNLO}~\cite{Ball:2014uwa} \\
$WZZ (5\ell1\nu)$ & \textsc{Sherpa 2.2.2} + \textsc{MEPS@NLO}~\cite{Hoeche:2011fd,Hoeche:2012yf,Hoeche:2009rj} & NLO & \textsc{Sherpa} and \textsc{NNPDF3.0 NNLO}~\cite{Ball:2014uwa} \\
$WWZ(4\ell2\nu)$ & \textsc{Sherpa 2.2.2} + \textsc{MEPS@NLO}~\cite{Hoeche:2011fd,Hoeche:2012yf,Hoeche:2009rj} & NLO  & \textsc{Sherpa} and \textsc{NNPDF3.0 NNLO}~\cite{Ball:2014uwa} \\
$Z$+jets & \textsc{Sherpa 2.2.1} + \textsc{MEPS@NLO}~\cite{Hoeche:2011fd,Hoeche:2012yf,Hoeche:2009rj} & NLO (0- and 2-jet), LO (3- and 4-jet) & \textsc{Sherpa} and \textsc{NNPDF3.0 NNLO}~\cite{Ball:2014uwa} \\
$t\bar{t}$ & \textsc{Powheg-Box v2} + \textsc{Pythia 8.230}~\cite{Alioli:2010xd,Nason:2004rx,Frixione:2007vw,Alioli:2010xd,Sjostrand:2014zea} & NLO+LO & A14 and (\textsc{NNPDF3.0 NLO} \& \textsc{NNPDF2.3 LO})~\cite{Lange:2001uf,Ball:2011mu,ATL-PHYS-PUB-2014-021}\\
$WZ\rightarrow 3\ell1\nu$ & \textsc{Powheg-Box v2} + \textsc{Pythia 8.230}~\cite{Alioli:2010xd,Nason:2004rx,Frixione:2007vw,Alioli:2010xd,Sjostrand:2014zea}   & NLO & \textsc{AZNLO} and (\textsc{CT10NLO} \& \textsc{CTEQ6L1})~\cite{STDM-2012-23,Lai:2010vv,Pumplin:2002vw} \\
\bottomrule
\end{tabular}}
\end{center}
\vspace{-0.3cm}
\caption{Summary of the event generators used for the simulated signal and background samples, including the accuracy of the matrix element and parton distribution functions (PDFs). Additionally, the table lists the set of tuned parameters used.} \label{samples}
\end{table}

%


%
\section{Object and event selection}
\label{sec:objects}

\subsection{Object reconstruction}
The event selection relies on reconstructing and identifying electrons, muons, jets and missing transverse momentum. The electron energy measurement is improved using a dynamic, topological calorimeter-cell clustering-based method. This method is particularly effective in cases where an electron radiates a bremsstrahlung photon. More information about this technique can be found in Ref.~\cite{EGAM-2018-01}. Electron candidates are identified as clusters of energy deposits in the calorimeter associated with ID tracks. The final track-cluster matching uses a Gaussian-sum filter (GSF)~\cite{ATLAS-CONF-2012-047}, accounting for bremsstrahlung energy losses. The electron transverse momentum ($p_{\textrm{T}}$) is calculated from the cluster energy and the track direction at the interaction point. The rejection of background noises, which consist of hadrons, non-prompt electrons mainly originating from photon conversions, and heavy flavour hadron decay, depends on the longitudinal and transverse shapes of the electromagnetic showers in the calorimeters, track-cluster matching, and properties of tracks in the ID. All this information, except for hits associated with the track, is used to create a likelihood discriminant. The selection criteria combine the likelihood with the number of track hits and define several working points (WP). Electrons with $p_{\textrm{T}}  > 4.5$~GeV and pseudorapidity range $|\eta| < 2.47$ are selected. The analysis uses a ``loose'' WP, with an efficiency of at least 90\% for electrons with $p_{\textrm{T}} > 30$~GeV~\cite{PERF-2017-01}.

Muon reconstruction is performed within $|\eta| < 2.7$~\cite{MUON-2018-03}, covering the entire detector. In the ID coverage range, muon reconstruction primarily applies a global fit of fully reconstructed tracks in the ID and the MS, resulting in combined muons. However, in the central region ($|\eta| < 0.1$), where the MS coverage is limited, alternative methods are employed for muon identification. Muons in this region can be identified either by matching a fully reconstructed ID track to an MS track segment (segment-tagged muons) or by associating with a calorimetric energy deposit consistent with that of a minimum-ionising particle (calorimeter-tagged muons). In these cases, the muon momentum is determined by the ID track alone. In the forward MS region ($2.5 < |\eta| < 2.7$) outside the ID coverage, standalone muons are accepted if they have hits in all three MS layers. Additionally, if forward ID tracks are available, they are combined with standalone MS tracks to form silicon-associated forward muons. Calorimeter-tagged muons are required to have $p_{\textrm{T}} > 15$~GeV, while all other muon candidates must have a minimum transverse momentum of 5~GeV. This analysis uses a ``loose'' muon identification WP that uses all muon types and has an efficiency of 98.5\%~\cite{MUON-2018-03}.

Fake leptons from processes such as $WZ$ and $Z$+jets are suppressed using impact-parameter and track- and calorimeter-based isolation requirements~\cite{PERF-2017-01,MUON-2018-03}. The transverse impact-parameter significance is the impact parameter calculated with respect to the measured beamline position in the transverse plane divided by its uncertainty, $|d_0|$/$\sigma_{d_0}$.  This significance is required to be less than 3 for muons and less than 5 for electrons. Leptons must be isolated using both track- and calorimeter-based information. The track-based discriminants take into account the scalar $p_{\textrm{T}}$ sum of all tracks in a cone of width $\Delta R$ = 0.3 for muons and 0.2 for electrons (excluding the lepton itself). The ratio of this sum to the lepton $p_{\textrm{T}}$ is required to be less than 0.15. The pile-up contributions are suppressed by requiring the cone tracks to originate from the primary vertex. To retain efficiency at higher $p_{\textrm{T}}$, the size of the track-isolation cone is reduced to 10~GeV/$p_{\textrm{T}}$ for $p_{\textrm{T}}$ above 33~GeV for muons and 50~GeV for electrons. A vertexing algorithm is used to fit the ID tracks of the $4\ell$ candidates under the assumption that they emanate from a common vertex. The resulting fit quality, represented by the $\chi^2/\textrm{ndof}$ value, provides good discrimination between signal and background. For the $4\mu$ channel, a cut of $\chi^2/\textrm{ndof}<9$ is applied, while a cut of $\chi^2/\textrm{ndof}<6$ is used for the other channels. This maintains a signal efficiency greater than 99\% for all channels.

The reconstruction of jets applies a particle-flow algorithm~\cite{PERF-2015-09}, which combines measurements from the tracker and the calorimeter. The first step is to remove those calorimeter hits that are associated with reconstructed charged particles. Then, the remaining calorimeter energy and tracks matched to the hard interaction are used to create particle-flow objects for the jet reconstruction. The anti-$k_t$ algorithm with a radius parameter of $R = 0.4 $ is used to reconstruct particle-flow jets~\cite{Cacciari:2008gp}. After the jets are reconstructed, pile-up is subtracted independently of the calibration~\cite{JETM-2018-05}. Jets are required to have $p_{\textrm{T}} > 30$~GeV and $|\eta| < 2.5$. A multivariate jet-vertex-tagger algorithm (JVT) is employed to reject low-momentum jets originating from pile-up events. The JVT considers the reconstructed tracks' properties in the event and applies to jets with $p_{\textrm{T}} < 60$~GeV and $|\eta| < 2.4$~\cite{PERF-2014-03,ATLAS-CONF-2014-018}.

Events containing $b$-hadrons ($b$-jets) are identified using a multivariate tagging algorithm ($b$-tagging)~\cite{FTAG-2018-01,ATL-PHYS-PUB-2017-013}. This algorithm assesses the likelihood that a jet originates from $b$-quark hadronisation by analysing track impact parameters and reconstructed secondary vertices. In simulated $t\overline{t}$ events, the jet identification algorithm exhibits an average efficiency of 77\% for identifying jets from $b$-quarks. Additionally, it achieves a rejection factor of approximately 30 for light-flavour quarks ($u$-, $d$-, and $s$-quarks), while the rejection factor for $c$-quarks and gluons is smaller, as reported in Ref.~\cite{FTAG-2018-01}. %

After object reconstruction, an overlap removal procedure is applied to all selected objects to remove ambiguities resulting from objects being reconstructed by several algorithms. The electron with the highest $p_{\textrm{T}}$ is kept when two electrons have overlapping energy deposits. If a reconstructed electron and muon share the same ID track, the electron is selected if the muon is calorimeter-tagged; otherwise, the muon is selected. Reconstructed jets which overlap with electrons (muons) in a cone of size $\Delta R = 0.2\,(0.1)$ are removed.

The magnitude of the missing transverse momentum is denoted by $E^{\textrm{miss}}_{\textrm{T}}$ and is calculated as the magnitude of the negative vectorial sum of the transverse momenta of calibrated leptons and jets. Also, a soft term is included in the calculation, which is constructed from all tracks originating from the primary vertex but not associated with any identified lepton or jet~\cite{ATLAS-CONF-2018-023}. Selected events must have at least one vertex with two associated tracks with $p_{\textrm{T}} > 500$ MeV, and the primary vertex is chosen to be the vertex with the largest $\sum p^2_{\textrm{T}}$.

\subsection{Event selection}\label{subsec:event}
This analysis classifies events into three channels based on the flavours of the selected four leptons: $4\mu$, $4e$, and $2\mu2e$. These channels are assigned based on the triggers activated in the event, which include single-lepton, dilepton, and trilepton triggers. The possible quadruplets in each channel of an event are created from two same flavour and opposite sign lepton pairs (SFOS). The $p_{\textrm{T}}$ thresholds of the three leading leptons in the quadruplet follow this order: 20~GeV, 15~GeV, and 10~GeV. The leading and the sub-leading lepton pairs in the quadruplet are defined to have an invariant mass closest and second closest to the $Z$ boson mass. The invariant mass of the leading and sub-leading lepton pairs are denoted $m_{Z_1}$ and $m_{Z_2}$, respectively. In the selected quadruplet, $m_{Z_1}$ must be within 50~GeV $<m_{Z_1}<106$~GeV, while $m_{Z_2}$ must be within 50~GeV $< m_{Z_2}<115$~GeV. The selected quadruplets must have their lepton pairs separated from each other by  $\Delta R > 0.1$. If any two SFOS leptons are detected with $m_{\ell\ell}$ below 5~GeV for $4\mu$ and $4e$ quadruplets, the quadruplet is excluded to reduce the contamination from $J/\psi$ mesons. In this analysis, the $Z+\Upsilon$ background contribution is negligible. If, at this point, several quadruplets of different channels are selected, only the quadruplet with the highest expected signal rate is retained, in the order: $4\mu$, $2e2\mu$, $4e$.

The resolution of the four lepton invariant mass system can be improved by recovering the radiative photon production in the $Z$ boson decay (final-state-radiation (FSR)) and by applying constraints on the $Z$ boson mass. In this analysis, the methodology presented in Refs.~\cite{HIGG-2013-20, HIGG-2016-19,HIGG-2018-09} for the $h\rightarrow ZZ$ analysis was used to account for FSR photons in the reconstruction of $Z$ bosons, and for applying constraints on the $Z$ boson mass. The $Z$-mass constraint is applied to both $Z$ candidates and improves the resolution of $m_{4\ell}$ by around 15\%. This analysis employs the combined $4\mu$, $4e$, and $2\mu2e$ channels, collectively called $4\ell$, to account for the limited statistics.


%
\section{Analysis strategy}
\label{sec:strategy}

This analysis searches for heavy resonances that decay to four leptons and missing transverse momentum or jets, where the four-lepton invariant mass is above 200~GeV. The main source of background in this kinematic region involves the leptonic decay of two $Z$ bosons. Among the various background processes, quark-antiquark annihilation ($q\overline{q}\rightarrow ZZ$) is the most significant, contributing 84.6\% of the expected background, while gluon-initiated production ($gg\rightarrow ZZ$) accounts for approximately 11.7\%. Other SM backgrounds, such as $q\overline{q}\rightarrow ZZ$ (EW), $t\bar{t}V$, $Z$+jets, $t\bar{t}$, $WZ$, and $VVV$, account for about 3.7\% of the total background events. These numbers are estimated from simulation after the four-lepton quadruplet selection discussed in Section~\ref{subsec:event}. In the following, this study models the SM backgrounds using an analytical function, illustrated in Section~\ref{sec:modelling}, where the normalisation is taken from the data.

\begin{table}[t]
\begin{center}
\begin{tabular}{|l|cccc|}
\hline
Signal region & \multicolumn{4}{c|}{$R\rightarrow SH\rightarrow 4\ell+E^{\textrm{miss}}_{\textrm{T}}$ and $A\rightarrow ZH\rightarrow 4\ell+X$}                                                                                                                             \\ \hline
SR1           & \multicolumn{1}{c|}{\multirow{3}{*}{$n_{b\textrm{-jets}}=0$}} & \multicolumn{1}{c|}{$n_{\textrm{jets}} = 0$}                     & \multicolumn{1}{c|}{$p^{4\ell}_{\textrm{T}}> 20$~GeV} & $E^{\textrm{miss}}_{\textrm{T}}$ significance $>$2.0        \\ \cline{1-1} \cline{3-5}
SR2           & \multicolumn{1}{c|}{}                                         & \multicolumn{1}{c|}{\multirow{2}{*}{$n_{\textrm{jets}} \geq 1$}} & \multicolumn{1}{c|}{$p^{4\ell}_{\textrm{T}}> 10$~GeV} & $E^{\textrm{miss}}_{\textrm{T}}$ significance $>3.5$        \\ \cline{1-1} \cline{4-5}
SR3           & \multicolumn{1}{c|}{}                                         & \multicolumn{1}{c|}{}                                            & \multicolumn{1}{c|}{$p^{4\ell}_{\textrm{T}}< 10$~GeV} & $2.5 < E^{\textrm{miss}}_{\textrm{T}}$ significance $<3.5 $ \\ \hline
& \multicolumn{4}{c|}{$A\rightarrow ZH\rightarrow 4\ell+X$}                                                                                                                                                                                                   \\ \hline
SR4           & \multicolumn{1}{c|}{\multirow{3}{*}{$n_{b\textrm{-jets}}=0$}} & \multicolumn{1}{c|}{\multirow{2}{*}{$n_{\textrm{jets}} \geq 2$}} & \multicolumn{2}{c|}{$|m_{jj} - m_{Z}| < 20$~GeV}                                                                             \\ \cline{1-1} \cline{4-5}
SR5           & \multicolumn{1}{c|}{}                                         & \multicolumn{1}{c|}{}                                            & \multicolumn{2}{c|}{$|m_{jj} - m_{Z}| > 20$~GeV}                                                                             \\ \cline{1-1} \cline{3-5}
SR6           & \multicolumn{1}{c|}{}                                         & \multicolumn{3}{c|}{$n_{\textrm{jets}} = 1$}                                                                                                                                              \\ \hline
SR7           & \multicolumn{4}{c|}{$n_{b\textrm{-jets}} \geq 1$}                                                                                                                                                                                                           \\ \hline
\end{tabular}
\end{center}
\vspace{-0.1cm}
\caption{Summary of the requirements for the signal regions (SR) defined in the search for the $R\rightarrow SH\rightarrow 4\ell+E^{\textrm{miss}}_{\textrm{T}}$ and $A\rightarrow ZH\rightarrow 4\ell+X$ signals for $m_{4\ell} > 200$~GeV.}\label{SRs}
\end{table}

A cut-based optimisation was used to increase the sensitivity of the $R\rightarrow SH\rightarrow 4\ell+E^{\textrm{miss}}_{\textrm{T}}$ and $A\rightarrow ZH\rightarrow 4\ell+X$ signals. This involves setting thresholds on various kinematic and topological variables, such as the number of jets ($n_{\textrm{jets}}$) and $b$-jets ($n_{b\textrm{-jets}}$), the transverse momentum of the four-lepton system ($p^{4\ell}_{\textrm{T}}$), and the $E^{\textrm{miss}}_{\textrm{T}}$ significance $\left(E^{\textrm{miss}}_{\textrm{T}}/\sqrt{\sum E_{\textrm{T}}}\right)$. The term $\sum E_{\textrm{T}}$ represents the scalar sum of the transverse energies of all reconstructed objects in the ATLAS detector. These variables exhibit kinematic differences between $R\rightarrow SH\rightarrow 4\ell+E^{\textrm{miss}}_{\textrm{T}}$ and $A\rightarrow ZH\rightarrow 4\ell+X$ signals and the background processes. The background composition varies depending on the $n_{\textrm{jets}}$ and $n_{b\textrm{-jets}}$. For the $p^{4\ell}_{\textrm{T}}$, most background processes peak around 20 GeV, with a tail stretch toward higher values. In contrast, the signal distribution peaks around 40 GeV. This difference in peak positions reflects the characteristic energy distribution of the signal compared to the background processes. Similarly, the $E^{\textrm{miss}}_{\textrm{T}}$ significance distribution also demonstrates distinguishable signal and background patterns. Background events peak at a $E^{\textrm{miss}}_{\textrm{T}}$ significance of around 1, indicating a relatively low significance of missing transverse energy. In contrast, the signal distribution peaks at a higher $E^{\textrm{miss}}_{\textrm{T}}$ significance of around 3, suggesting a stronger presence of genuine missing transverse energy associated with the signal process. Therefore, events with no $b$-jets are divided into two categories based on the $n_{\textrm{jets}}$: those with no jets ($n_{\textrm{jets}}=0$) and those with at least one jet ($n_{\textrm{jets}} \geq 1$). A two-dimensional scan is applied twice to each group using $p^{4\ell}_{\textrm{T}}$ and $E^{\textrm{miss}}_{\textrm{T}}$ significance to find the optimal cut between signal and background processes. The first scan determines the optimal threshold values for these variables corresponding to the highest value of the ratio of signal yields to the square root of background yields. After finding the optimal points, the events above these optimal thresholds are removed, and then a second scan is performed on the remaining events. This process results in specific requirements on the variables that define the signal regions (SR). Therefore, events in the SR are required to satisfy the following requirements:
\begin{itemize}
\item  $n_{b\textrm{-jets}}=0$ and $n_{\textrm{jets}}=0$ with $p^{4\ell}_{\textrm{T}}>20$~GeV and $E^{\textrm{miss}}_{\textrm{T}}$ significance $>$ 2.0 (SR1)
\item  $n_{b\textrm{-jets}}=0$ and $n_{\textrm{jets}}\geq 1$ with $p^{4\ell}_{\textrm{T}}> 10$~GeV and $E^{\textrm{miss}}_{\textrm{T}}$ significance $>$ 3.5 (SR2)
\item  $n_{b\textrm{-jets}}=0$ and $n_{\textrm{jets}}\geq 1$ with $p^{4\ell}_{\textrm{T}}<10$~GeV and $2.5<E^{\textrm{miss}}_{\textrm{T}}$ significance $<3.5$ (SR3)
\end{itemize}
The SRs mentioned above are used for both the $R\rightarrow SH\rightarrow 4\ell+E^{\textrm{miss}}_{\textrm{T}}$ and the $A\rightarrow ZH\rightarrow 4\ell+X$ models.

Jets in the $A\rightarrow ZH\rightarrow 4\ell+X$ signal can arise from initial state radiation and the decay of the $Z$ boson. Therefore, it is crucial to accommodate scenarios with varying jet multiplicities and recover the $b$-jet events left from the abovementioned categorisation. Additionally, the invariant mass of the two leading jets ($m_{jj}$) shows a significant difference between the $A\rightarrow ZH\rightarrow 4\ell+X$ signal and backgrounds at $m_{jj} > 300$ GeV. Given these differences, more categorisations were designed for the $A\rightarrow ZH\rightarrow 4\ell+X$ signal. This expanded categorisation comprises four additional SRs to capture distinct jet-related features:
\begin{itemize}
\item $n_{b\textrm{-jets}}=0$, $n_{\textrm{jets}}\geq 2$, $|m_{jj} - m_{Z}| < 20$~GeV (SR4)
\item $n_{b\textrm{-jets}}=0$, $n_{\textrm{jets}}\geq 2$, $|m_{jj} - m_{Z}| > 20$~GeV (SR5)
\item $n_{b\textrm{-jets}}=0$, $n_{\textrm{jets}}= 1$ (SR6)
\item $n_{b\textrm{-jets}}\geq 1$ (SR7)
\end{itemize}
These SRs are only considered if the events are not yet classified in SR1-3. SR6 is motivated by the need to capture signal events in which only one jet is detected. A single jet event occurs if the produced jet has very low momentum and falls outside the detector acceptance. SR7 is designed to include events with at least one $b$-jet, primarily due to the significant branching ratio of the $Z$ boson into $b$ quarks. Table~\ref{SRs} provides a summary of these seven SRs used in the analysis.


%
\section{Signal and background modelling}
\label{sec:modelling}

Simulation is used to model and then parameterise the constructed $m_{4\ell}$ distribution for the SM backgrounds. Meanwhile, the signal shape is taken directly from the simulation, as illustrated below.

\subsection{Signal model}
Section~\ref{sec:data} discussed the generation of signal mass points for $R\rightarrow SH\rightarrow 4\ell+E^{\textrm{miss}}_{\textrm{T}}$ and $A\rightarrow ZH\rightarrow 4\ell+X$ signals. Seventy-seven mass points were produced for the $R\rightarrow SH\rightarrow 4\ell+E^{\textrm{miss}}_{\textrm{T}}$ signal, while seventy-two were generated for the $A\rightarrow ZH\rightarrow 4\ell+X$ signal. To cover a broader mass range, a linear interpolation method described in Ref.~\cite{Read:1999kh} is used to obtain signal shapes between the generated mass planes in either of the $(m_{R}, m_H)$ or $(m_{A}, m_H)$ masses. Interpolating the signal involves a two-step process due to its dependence on either $(m_{R}, m_{H})$ or $(m_{A}, m_{H})$ masses. In the first step, the $H$ mass remains constant while interpolating the $(m_{R} - m_{H})$ or $(m_{A} - m_{H})$ mass parameters or interpolating the $R$ or $A$ mass. Subsequently, the interpolated signals obtained in the previous step serve as input for a second interpolation. In this stage, the $R$ or $A$ mass is fixed, while the $H$ mass varies in steps of 10~GeV.

\begin{figure}[!t]
\begin{center}
\subfloat[]{\includegraphics[width=0.45\linewidth]{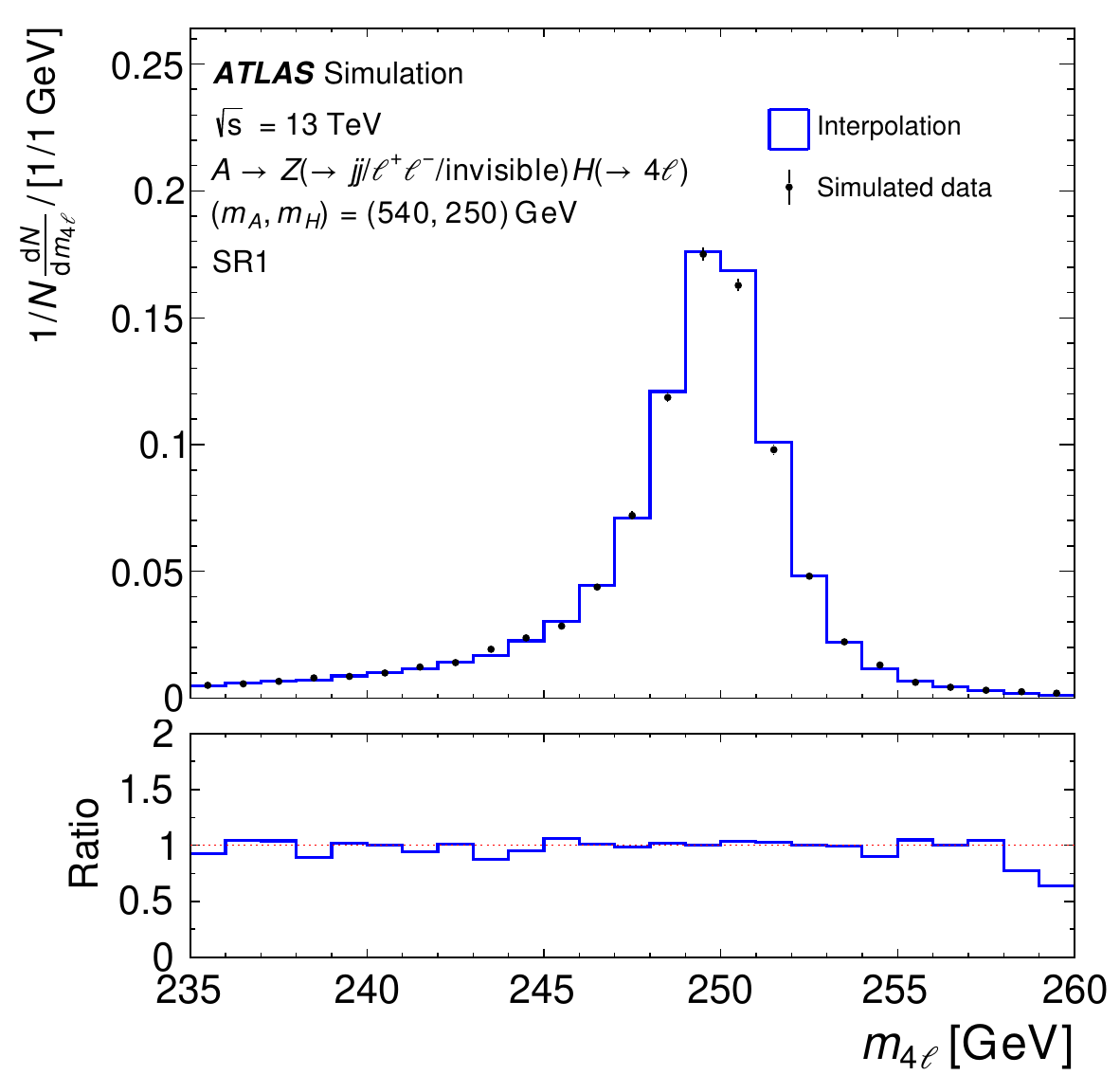}\label{HMT0cjet_H4l}}\;\;
\subfloat[]{\includegraphics[width=0.45\linewidth]{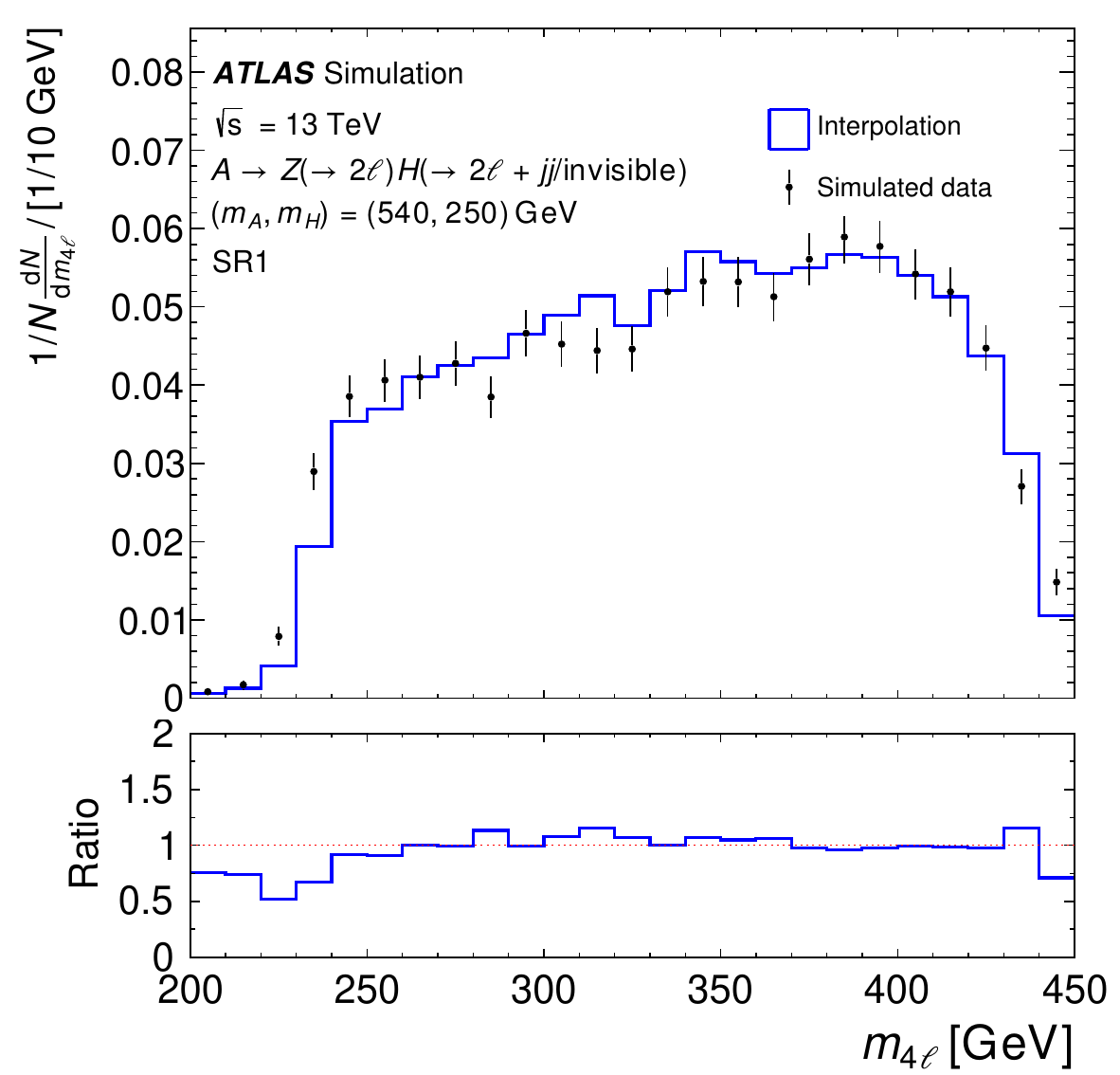}\label{HMT0cjet_H2l}}\\
\subfloat[]{\includegraphics[width=0.45\linewidth]{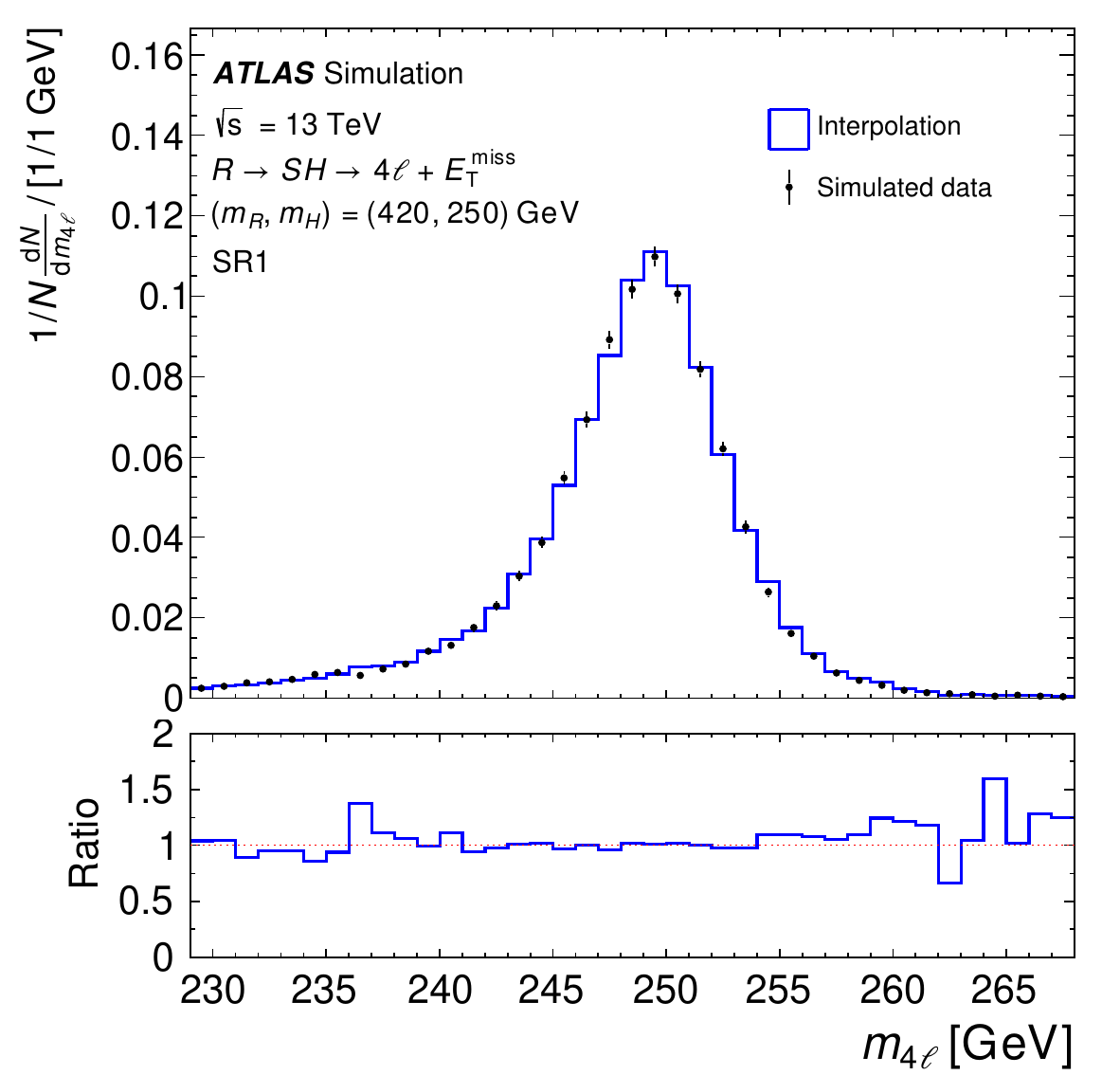}\label{HMT0cjet_RSH}}
\end{center}
\vspace{-0.5cm}
\caption{The $m_{4\ell}$ distributions of the interpolated (blue) and simulated (black-filled point) signals for the \protect\subref{HMT0cjet_H4l} $A\rightarrow Z(\rightarrow jj/\ell^+\ell^-/\textrm{invisible})H(\rightarrow 4\ell)$ and \protect\subref{HMT0cjet_H2l} $A\rightarrow Z(\rightarrow \ell^+\ell^-)H(\rightarrow 2\ell+jj/\textrm{invisible})$ signals with the $(m_{A}, m_{H}) = (540, 250)$~GeV mass point and the \protect\subref{HMT0cjet_RSH} $R\rightarrow SH\rightarrow 4\ell+E^{\textrm{miss}}_{\textrm{T}}$ signal with $(m_{R}, m_{H}) = (420, 250)$~GeV and $m_S = 160$~GeV. The lower panels show the ratio between the interpolated and simulated histograms. The plots are only shown in SR1 for illustrative purposes.}\label{modelling}
\end{figure}

The interpolation was validated by comparing simulated and interpolated signal distributions at a few mass points, as shown in Figure~\ref{modelling} for the $A\rightarrow ZH\rightarrow 4\ell+X$ $R\rightarrow SH\rightarrow 4\ell+E^{\textrm{miss}}_{\textrm{T}}$ signals. Some disagreement at the tails of the signals is observed and considered as potential systematic uncertainty. This systematic uncertainty was shown to have a negligible impact on the results and was omitted.

\subsection{Background model}\label{bkg_modelling}
The $m_{4\ell}$ shape of the backgrounds is obtained from simulation using a parameterised empirical function to decrease statistical uncertainties originating from the limited number of simulated events. Four background templates are used: $q\overline{q}\rightarrow ZZ$, $gg\rightarrow ZZ$, $VVV$, and other background processes. The other backgrounds contain processes such as $q\overline{q}\rightarrow ZZ$ (EW), $t\overline{t}V$, $t\overline{t}$, $Z$+jets and $WZ$. Each of the background templates is fitted with an analytical function for $m_{4\ell}$ between 200--1200~GeV, as follows:
\begin{figure}[tbp]
\begin{center}
\subfloat[]{\includegraphics[width=0.45\linewidth]{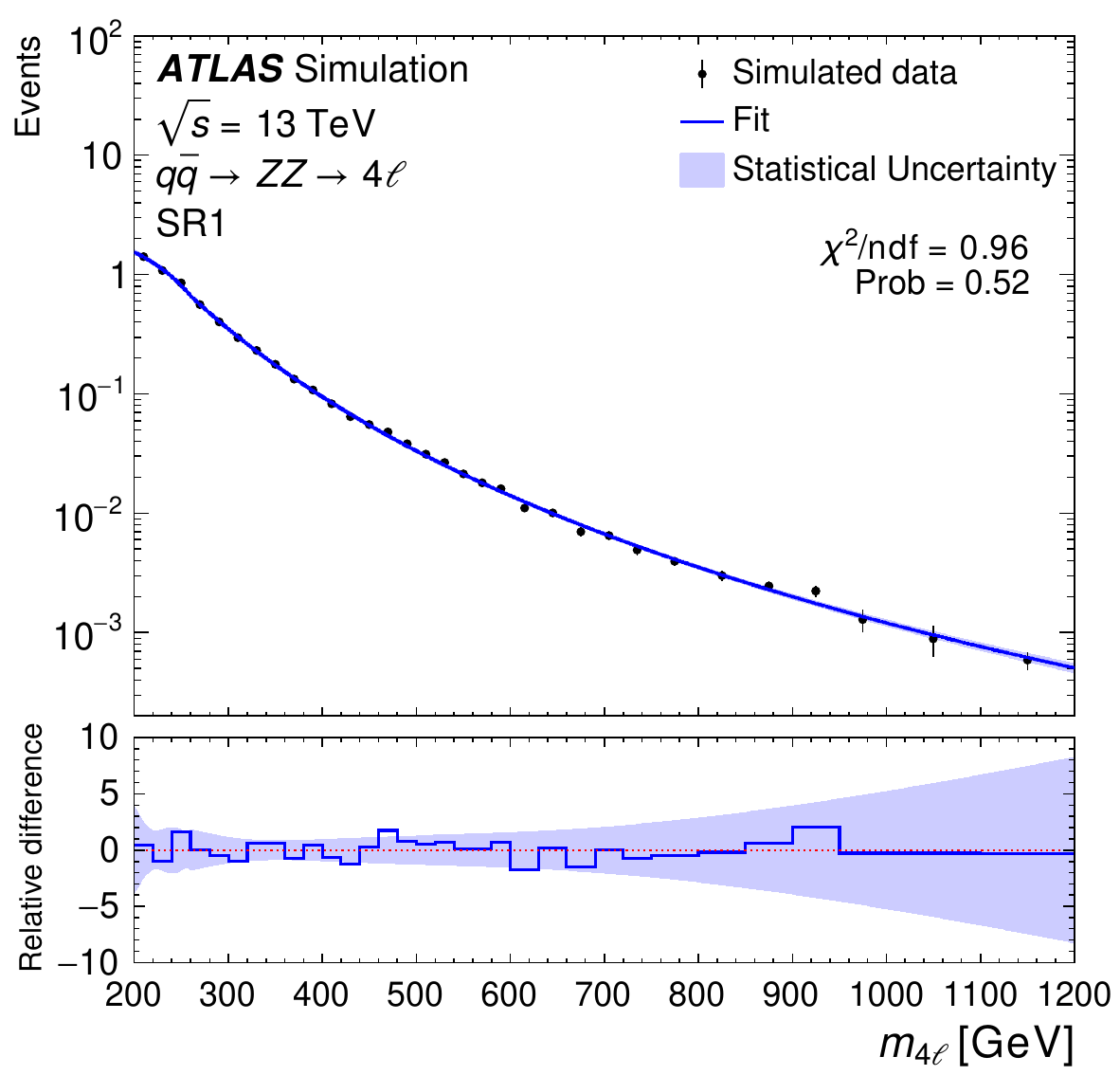}\label{qqZZ_SR1}}\;\;
\subfloat[]{\includegraphics[width=0.45\linewidth]{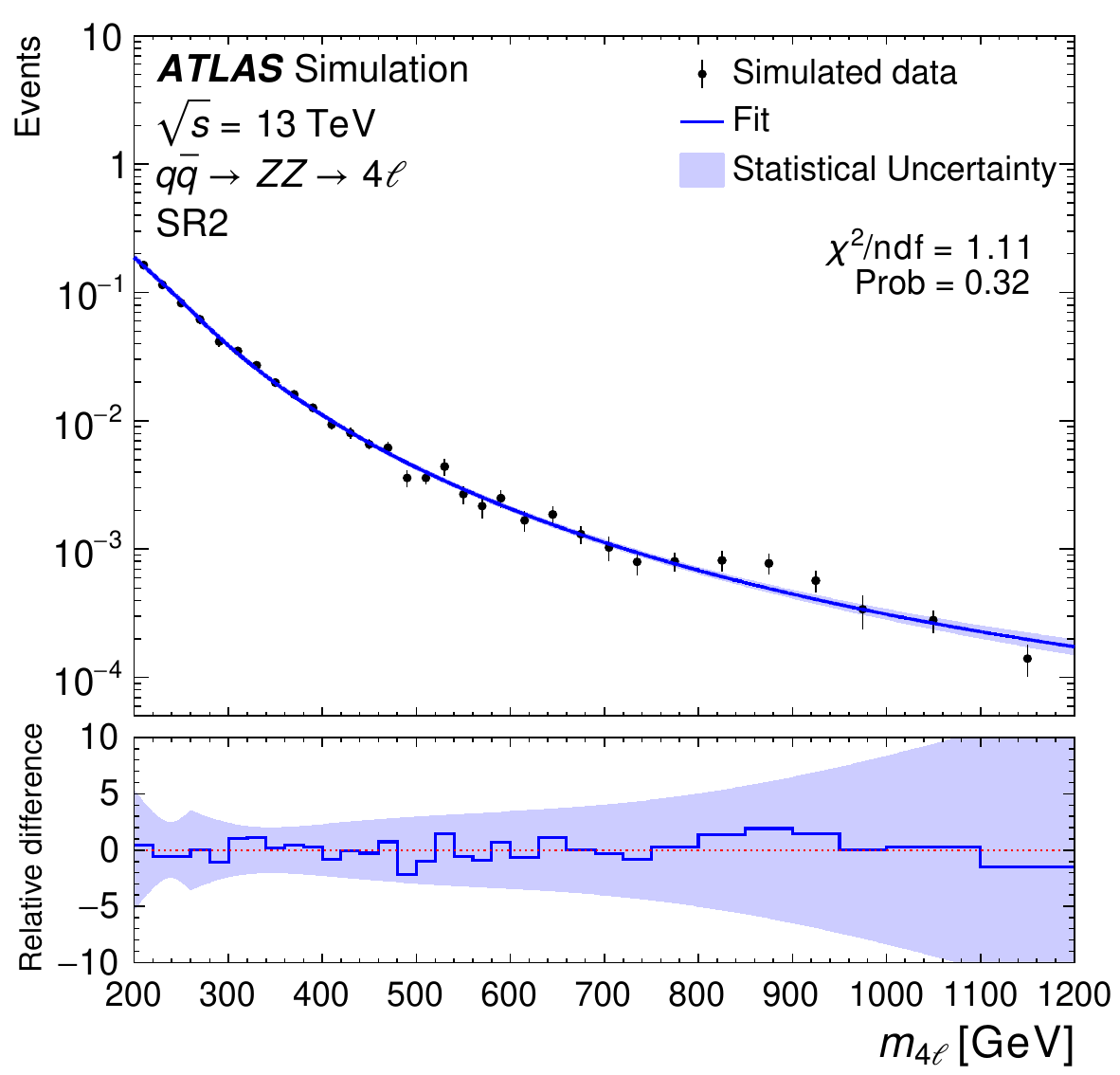}\label{qqZZ_SR2}}
\end{center}
\vspace{-0.5cm}
\caption{The $m_{4\ell}$ distributions of the analytical function fit (blue) to the simulated $q\overline{q}\rightarrow ZZ$ background (black-filled point) in signal regions~\protect\subref{qqZZ_SR1} SR1 and~\protect\subref{qqZZ_SR2} SR2. The lower panels show the relative difference between the fitted function and simulated histograms. The overlaid light blue band indicates the statistical uncertainty of the fit function.}\label{bkg_mod}
\end{figure}
\begin{eqnarray}
f (m_{4\ell}) = H(m_{0} - m_{4\ell})f_{1} (m_{4\ell})\,C_1 + H(m_{4\ell} - m_{0})f_{2} (m_{4\ell})\,C_2,\label{eq:BackgroundModel1}
\end{eqnarray}
where:
\begin{eqnarray}
f_{1} (m_{4\ell}) &=& \frac{a_1\cdot m_{4\ell} + a_2\cdot m^2_{4\ell}}{1+\exp\left(\frac{m_{4\ell} - a_1}{a_3}\right)}, \\
f_{2} (m_{4\ell}) &=& \left(1 - \frac{m_{4\ell}}{n_{C}}\right)^{b_{1}}\cdot \left(\frac{m_{4\ell}}{n_{C}}\right)^{\left(b_{2} + b_{3}\cdot \ln(\frac{m_{4\ell}}{n_{C}})\right)},
\end{eqnarray}
\begin{equation}
C_1 = \frac{1}{f_{1} (m_0)},\qquad C_2 = \frac{1}{f_{2} (m_0)}.\nonumber
\end{equation}
The functional form $f_{1}$ models the $ZZ$ threshold around $2\cdot m_{Z}$, and $f_{2}$ describes the high mass tail. The transition between the $f_{1}$ and $f_{2}$ functions is performed by the Heaviside step function $H(x)$ around $m_{0}$, where $m_{0}$ is fixed to 260~GeV, 240~GeV, 250~GeV and 230~GeV for the $q\overline{q}\rightarrow ZZ$, $gg\rightarrow ZZ$, $VVV$, and other backgrounds, respectively. The transition point is determined by the smoothness of the functional form. A constant $n_C=13$ TeV scales $m_{4\ell}$ in the high mass region. $C_1$ and $C_2$ ensure the continuity of the function around the point $m_0$, where $f_1$ and $f_2$ are joined together. The parameters $a_i$ and $b_i$ are determined by fitting the $m_{4\ell}$ distribution in simulation for each specific SR. Overall, all regions achieved a good fit quality with the empirical function, assessed by the chi-squared values, which are approximately equal to unity for all signal regions. As an example, Figure~\ref{bkg_mod} demonstrates the fit of the analytical function given by Equation~\ref{eq:BackgroundModel1} to the dominant $q\overline{q}\rightarrow ZZ$ background process in SR1 and SR2. The integral error calculated from the fit function is considered a source of systematic uncertainty. It is treated as a nuisance parameter and propagated as a shape systematic uncertainty to the final fit.
%


%
\section{Systematic uncertainties}
\label{sec:systematic}

This section discusses the sources of systematic uncertainty considered in this analysis, including experimental and theoretical uncertainties for the signal and background processes. Statistical uncertainties arising from the MC event generation are also considered. Additionally, uncertainties introduced by the background parameterisation are considered and propagated to the final fit. The systematic uncertainties are evaluated using the invariant mass of the four lepton system as a discriminant. Each systematic uncertainty component is treated as a nuisance parameter and profiled in the final statistical analysis.
\subsection{Experimental uncertainties}
Experimental uncertainties, such as object reconstruction and identification, trigger efficiencies, energy scale and resolution, are calculated for signal and background processes. The systematic experimental uncertainties are evaluated by computing the ratio between the integral of the $m_{4\ell}$ distribution with the relevant nuisance parameter weight modified by one standard deviation and the integral of the $m_{4\ell}$ distribution using the nominal weight. Signal samples were divided into three categories depending on the size of the energy gap ($m_A-m_H$ or $m_R-m_H$) to simplify estimating the experimental systematic uncertainties for the signals. The gaps include samples with low ($<500$~GeV), medium ($>500$~GeV and $<700$~GeV) and high ($>1300$~GeV) energy gaps. These ranges are for the $R\rightarrow SH\rightarrow 4\ell+E^{\textrm{miss}}_{\textrm{T}}$ process, and they are slightly different for the $A\rightarrow ZH\rightarrow4\ell+X$ process. The systematic uncertainties are evaluated for three signal mass points from each energy gap for both signals. Each result is combined by calculating the mean of the three samples per energy gap. Only the effect on the absolute normalisation was found to be relevant for the signal. For the $VVV$ and other backgrounds, the absolute normalisation and the shape variation were included in the fit. In contrast, only the shape variation for the $q\bar{q}\rightarrow ZZ$ and $gg\rightarrow ZZ$ was included as systematic uncertainty in the fit.

The lepton identification, isolation and reconstruction efficiency, energy/momentum scale, and resolution are derived from data using $J/\psi\rightarrow \ell\ell$ and $Z\rightarrow \ell\ell $ events. The uncertainties in the lepton efficiency are calculated following the method presented in Refs.~\cite{PERF-2015-10,PERF-2017-01} for muons and electrons. Generally, their effect on the signal yields is less than 1\% for the $R\rightarrow SH\rightarrow 4\ell+E^{\textrm{miss}}_{\textrm{T}}$ signal, and up to 3.6\% for the $A\rightarrow ZH\rightarrow4\ell+X$ signal. These uncertainties are no more than 1.5\% for the $VVV$ background yields and at most 1.7\% for the other background processes. The uncertainties in the jet energy scale and resolution come from sources such as uncertainties in the absolute and relative ``in situ'' calibration and the correction for pile-up~\cite{JETM-2018-05}. In total, uncertainties due to the jet energy scale and energy resolution are estimated to a maximum of 4.3\% for the $R\rightarrow SH\rightarrow 4\ell+E^{\textrm{miss}}_{\textrm{T}}$ signal, 4.7\% for the $A\rightarrow ZH\rightarrow4\ell+X$ signal, 7.0\% for the $VVV$ background and 10.4\% for the other background processes. In addition, flavour tagging uncertainties are calculated by altering the uncertainty of the tagged jet flavour efficiency by $\pm 1\sigma$. The total estimated flavour tagging uncertainties are as high as 4.3$\%$ for the $R\rightarrow SH\rightarrow 4\ell+E^{\textrm{miss}}_{\textrm{T}}$ and  $A\rightarrow ZH\rightarrow4\ell+X$ signals, 8.9\% for the $VVV$ background and 8.5\% for the other background processes. The uncertainty in the $E^{\textrm{miss}}_{\textrm{T}}$ is influenced by uncertainties in the energy scales of both leptons and jets~\cite{ATLAS:2020atr}. Furthermore, there are contributions to the $E^{\textrm{miss}}_{\textrm{T}}$ uncertainty from the momentum scale and resolution of tracks that are not associated with any identified lepton or jet. The overall uncertainty in $E^{\textrm{miss}}_{\textrm{T}}$ due to scale and resolution for the $R\rightarrow SH\rightarrow 4\ell+E^{\textrm{miss}}_{\textrm{T}}$ signal and $VVV$ background is less than 1\%. For the $A\rightarrow ZH\rightarrow4\ell+X$ signal, these uncertainties are up to 4.7\% and 6.2\% for the scale and resolution, respectively, while for the other background processes, it is at most 6.1\% due to the momentum scale and 9.7\% due to the resolution. These experimental systematic uncertainties have a negligible impact on the shape, so only normalisation uncertainties are considered.

The uncertainty in the combined 2015--2018 integrated luminosity is 1.7\%~\cite{ATLAS-CONF-2019-021}, obtained using the LUCID-2 detector~\cite{LUCID2} for the primary luminosity measurements. In addition, a systematic uncertainty from re-scaling the simulated pile-up to the data is considered. It is estimated to be up to 2.5\% for the $R\rightarrow SH\rightarrow 4\ell+E^{\textrm{miss}}_{\textrm{T}}$ signal, less than 1\% for the $A\rightarrow ZH\rightarrow4\ell+X$ signal, and about 3.0\% for the $VVV$ and other background processes.

In Section~\ref{sec:stat_model}, a common normalisation factor is introduced to scale the event yields for both $q\overline{q}\rightarrow ZZ$ and $gg\rightarrow ZZ$ backgrounds while keeping the $VVV$ and other backgrounds fixed to their SM predictions. Subsequently, the relative acceptance difference is calculated between the $q\overline{q}\rightarrow ZZ$ and $gg\rightarrow ZZ$ backgrounds for each source of the experimental systematic uncertainties. These experimental systematic uncertainties are then propagated to the $gg\rightarrow ZZ$ background. Experimental systematic uncertainties, which affect the shape of the $m_{4\ell}$ distribution of the $q\overline{q}\rightarrow ZZ$ and $gg\rightarrow ZZ$ backgrounds, are considered in the analysis as shape systematic uncertainties.

\subsection{Theoretical uncertainties}
This analysis considers the uncertainties of the PDF, missing high-order QCD correction, initial- and final-state radiation (ISR/FSR), and parton showering (PS) and hadronisation. The theoretical scale uncertainties on the signal samples are assessed by varying the factorisation and renormalisation scales up and down from their nominal values by a factor of two in each SR. The largest effect of the scale uncertainties is taken as the systematic uncertainty. The PDF uncertainty is estimated using the \textsc{SysCalc}~\cite{kalogeropoulos2018syscalc} package in conjunction with \textsc{Madgraph5\_aMC@NLO 2.9.5}. The uncertainty is determined by considering how the envelope of variations among alternative PDFs differs from the internal PDF error sets (\textsc{NNPDF2.3 PDF}). The calculation follows the PDF4LHC recommendations provided in Ref.~\cite{Butterworth:2015oua}. Uncertainties related to the ISR/FSR, PS and hadronisation are evaluated by employing variations in tune eigenvectors in \textsc{Pythia} and comparing \textsc{Herwig 7.1.3}~\cite{Bellm:2015jjp} with \textsc{Pythia 8.230} for the signal samples. The impact of the theoretical uncertainties on the signal acceptance is about 1\%. The theoretical systematic uncertainties on the SM backgrounds are included as a shape systematic on the $m_{4\ell}$ distribution and summarised below.

\subsection{Systematic uncertainties on background shapes}
To account for any possible biases introduced by the analytical function discussed in Section~\ref{bkg_modelling}, uncertainties stemming from the background parameterisation are considered. These uncertainties are calculated by evaluating the error in the fit parameters while also taking into account the errors in the parameters resulting from the fit. The errors are estimated using the correlation values obtained from the fit.

Incorporating PDF and QCD scale uncertainties into the analysis is motivated by their impact on the shape of the $m_{4\ell}$ distributions for the $q\overline{q}\rightarrow ZZ$ and $gg\rightarrow ZZ$ processes. The QCD scale uncertainty is estimated to be up to 10\% for the $q\overline{q}\rightarrow ZZ$ and $gg\rightarrow ZZ$ backgrounds. The PDF uncertainty is less than 5\% (1\%) in the low mass region and about 25\% (30\%) in the high mass region for the $q\overline{q}\rightarrow ZZ$ ($gg\rightarrow ZZ$) background. A missing higher-order QCD correction uncertainty on the $m_{4\ell}$ shape is considered for the $q\overline{q}\rightarrow ZZ$ background, which is estimated to be less than 10\%. The parton shower uncertainty is evaluated by varying parameters in the parton shower tunes, such as the CKKW and QSF settings, and by using different showering options.


%
\section{Statistical procedures}
\label{sec:stat_model}

The invariant mass of the four leptons is used as a discriminant to examine the null and alternative hypotheses using the profile likelihood ratio technique \cite{Cowan:2010js}.  The null hypothesis corresponds to smoothed backgrounds that fall from the low mass range to the higher mass range of the $m_{4\ell}$ distribution.  In contrast, the alternative hypothesis incorporates a signal structure around the $H$ mass. The signal and background contributions in the $m_{4\ell}$ distribution are extracted via a binned maximum-likelihood fit of the signal-plus background hypotheses to extract any indication for new physics. The profile likelihood function is defined as the probability of observing $n$ events. It is calculated as the product of the sum of weighted signal and background events, as shown below:
\begin{equation}{\cal L } (m^n_{4\ell}|\sigma(gg\rightarrow A/R), \vec{\theta}) =  \prod\limits^{n_r}_{r=\textrm{SRs}} \prod\limits^{n_i}_{i=\textrm{bin}} \textrm{Poisson}\left(n_{r, i}\left|\right. s_{r, i} + \sum_{b} b_{r, i}(\vec{\theta})\right)\times \prod\limits_{i} G_i(0|\vec{\theta},1),
\end{equation}
where $\sigma(gg\rightarrow A/R)$ is the parameter of interest for the $A\rightarrow ZH\rightarrow 4\ell+X$ or $R\rightarrow SH\rightarrow 4\ell+E^{\textrm{miss}}_{\textrm{T}}$ signal. The expected signal and background yields in each $m_{4\ell}$ distribution bin are represented by $s$ and $b$, respectively. The expected signal yield $s$  is calculated by:
\begin{equation}
s = \sigma(gg\rightarrow A/R)\times\mathcal{B}(A/R\rightarrow ZH/SH)\times\mathcal{B}(H\rightarrow ZZ)\times AC\times \int L~\textrm{d}t,
\end{equation}
where $AC$ is the acceptance times efficiency, and $\int L~\textrm{d}t = 139$ fb$^{-1}$ is the integrated luminosity of the data. A collection of nuisance parameters is introduced to describe how systematic uncertainties influence the predicted number of signal and background events and the shape of the PDFs. These parameters are constrained to their nominal values within the calculated uncertainties using Gaussian constraints by $G(\vec{\theta})$, in which $\vec{\theta}$ is a vector containing the nuisance parameters. In the likelihood fit function, SRs are indicated in the product by the index ``$r$''.  In each SR, the normalisation for the $q\overline{q}\rightarrow ZZ$ and $gg\rightarrow ZZ$ processes (ZZ background) are calculated by a likelihood fit to the data. The $ZZ$ background normalisation is denoted by the $\mu^{ZZ}_{\textrm{norm}}$ parameter. The benefit of taking the $ZZ$ background normalisation from the data is the reduction of the background dependence on the theoretical systematic uncertainties.


%
\section{Results}
\label{sec:result}

\begin{table}[b]
\sisetup{separate-uncertainty}
\begin{center}
\resizebox{\textwidth}{!}{%
\begin{tabular}{lccccccc}
\toprule \noalign{\smallskip}
& SR1 & SR2 & SR3 & SR4 & SR5 & SR6 & SR7 \\
\noalign{\smallskip}\midrule\noalign{\smallskip}
$q\overline{q}\rightarrow ZZ$&    \numRF{132.153}{3} $\pm$ \numRF{11.9826}{2} &
\numRF{16.8413}{2} $\pm$ \numRF{5.99228}{1} & \numRF{41.9034}{2} $\pm$ \numRF{6.83052}{1} & \numRF{39.7624}{2} $\pm$ \numRF{6.5614 }{1} & \numRF{156.051}{3} $\pm$ \numRF{12.7315}{2} & \numRF{549.113}{3} $\pm$ \numRF{69.5908}{1} & \numRF{86.2806}{2} $\pm$ \numRF{11.4287}{2} \\
$gg\rightarrow ZZ$           &    \numRF{31.9506}{2} $\pm$ \numRF{5.84207}{1} &
\numRF{3.21537}{2} $\pm$ \numRF{3.09636}{2} & \numRF{8.33268}{2} $\pm$ \numRF{3.05593}{2} & \numRF{6.61435}{2} $\pm$ \numRF{2.77647}{2} & \numRF{22.1306}{3} $\pm$ \numRF{3.65677}{2} & \numRF{102.011}{3} $\pm$ \numRF{71.4883}{1} & \numRF{9.39615}{2} $\pm$ \numRF{4.18338}{2} \\
$VVV$                        &    \numRF{7.4555  }{2} $\pm$ \numRF{0.393605 }{1} &
\numRF{4.77249 }{3} $\pm$ \numRF{0.335703 }{2} & \numRF{1.49396 }{3} $\pm$ \numRF{0.192419 }{2} & \numRF{0.185587}{2} $\pm$ \numRF{0.0372482}{1} & \numRF{1.07855 }{3} $\pm$ \numRF{0.106454 }{2} & \numRF{1.58723 }{3} $\pm$ \numRF{0.127188 }{2} & \numRF{0.501656}{2} $\pm$ \numRF{0.0709402}{1} \\
Other backgrounds            &    \numRF{5.49118}{2} $\pm$ \numRF{0.931501}{1} &
\numRF{7.09757}{2} $\pm$ \numRF{1.13116 }{2} & \numRF{3.56105}{2} $\pm$ \numRF{0.698769}{1} & \numRF{2.47406}{3} $\pm$ \numRF{0.302861}{2} & \numRF{17.5059}{3} $\pm$ \numRF{0.839004}{1} & \numRF{11.2515}{3} $\pm$ \numRF{0.765874}{1} & \numRF{37.6497}{3} $\pm$ \numRF{2.31744 }{2} \\
\midrule
Total background             &    \numRF{177.051}{3} $\pm$ \numRF{13.2948}{2} &
\numRF{31.9267}{2} $\pm$ \numRF{5.57551}{1} & \numRF{55.2911}{2} $\pm$ \numRF{7.40744}{1} & \numRF{49.0364}{2} $\pm$ \numRF{6.99573}{1} & \numRF{196.766}{3} $\pm$ \numRF{14.0283}{2} & \numRF{663.963}{3} $\pm$ \numRF{26.2931}{2} & \numRF{133.828}{3} $\pm$ \numRF{11.5854}{2} \\
\midrule
Observed                     &    177 &
32  &
55  &
49  &
197 &
664 &
135 \\
\midrule
$\mu^{ZZ}_{\textrm{norm}}$   & \numRF{1.14919 }{3} $\pm$ \numRF{0.148108 }{2} &
\numRF{1.2934  }{2} $\pm$ \numRF{0.594844 }{1} & \numRF{0.958844}{2} $\pm$ \numRF{0.217635 }{2} & \numRF{0.903619}{2} $\pm$ \numRF{0.170268 }{2} & \numRF{0.796767}{2} $\pm$ \numRF{0.0842455}{1} &
\numRF{1.02096 }{3} $\pm$ \numRF{0.134683 }{2} &
\numRF{1.42805 }{2} $\pm$ \numRF{0.330747 }{1} \\
\bottomrule
\end{tabular}}
\vspace{0.2cm}
\caption{Observed and expected post-fit event yields for $m_{4\ell}>200$~GeV with their uncertainties. The expected yields and their uncertainties are obtained from a simultaneous fit to data under the background-only hypothesis on all seven signal regions discussed in Section~\ref{sec:strategy}. The $\mu^{ZZ}_{\textrm{norm}}$ is the normalisation factor for the $q\overline{q}\rightarrow ZZ$ and $gg\rightarrow ZZ$ backgrounds. The other backgrounds include $q\overline{q}\rightarrow ZZ$ (EW), $t\overline{t}V$, $t\overline{t}$, $Z$+jets and $WZ$ processes. These backgrounds and the $VVV$ background process were fixed to their SM prediction.}\label{tab:yields}
\end{center}
\end{table}

Table~\ref{tab:yields} displays the yields in each of the seven SRs described in Section~\ref{sec:strategy}. The observed and expected numbers of events in each SR were obtained using a simultaneous binned maximum-likelihood fit, assuming the background-only hypothesis. Signal+background binned maximum-likelihood fits were also performed on the $m_{4\ell}$ distribution to search for potential excesses beyond the expected SM backgrounds. This fit model is based on the statistical framework discussed in Section~\ref{sec:stat_model}. The $m_R$ ($m_A$) range 390--1300 (320--1300)~GeV and the $m_H$ range 220--1000~GeV were scanned for the $R\rightarrow SH\rightarrow 4\ell+E^{\textrm{miss}}_{\textrm{T}}$ ($A\rightarrow ZH\rightarrow 4\ell+X$) search. The fit was performed in steps of 10~GeV, in which 4187 ($m_R, m_H$) and 4740 ($m_A, m_H$) mass points in NWA for each SR were tested for the $R\rightarrow SH\rightarrow 4\ell+E^{\textrm{miss}}_{\textrm{T}}$ and $A\rightarrow ZH\rightarrow 4\ell+X$ processes, respectively. The resulting test statistics were used to construct $p$-values and significance estimates, which were then used to evaluate the compatibility of the data with the background-only hypothesis and the presence of new physics in the data.

\begin{figure}[!t]
\begin{center}
\subfloat[]{\includegraphics[width=0.45\linewidth]{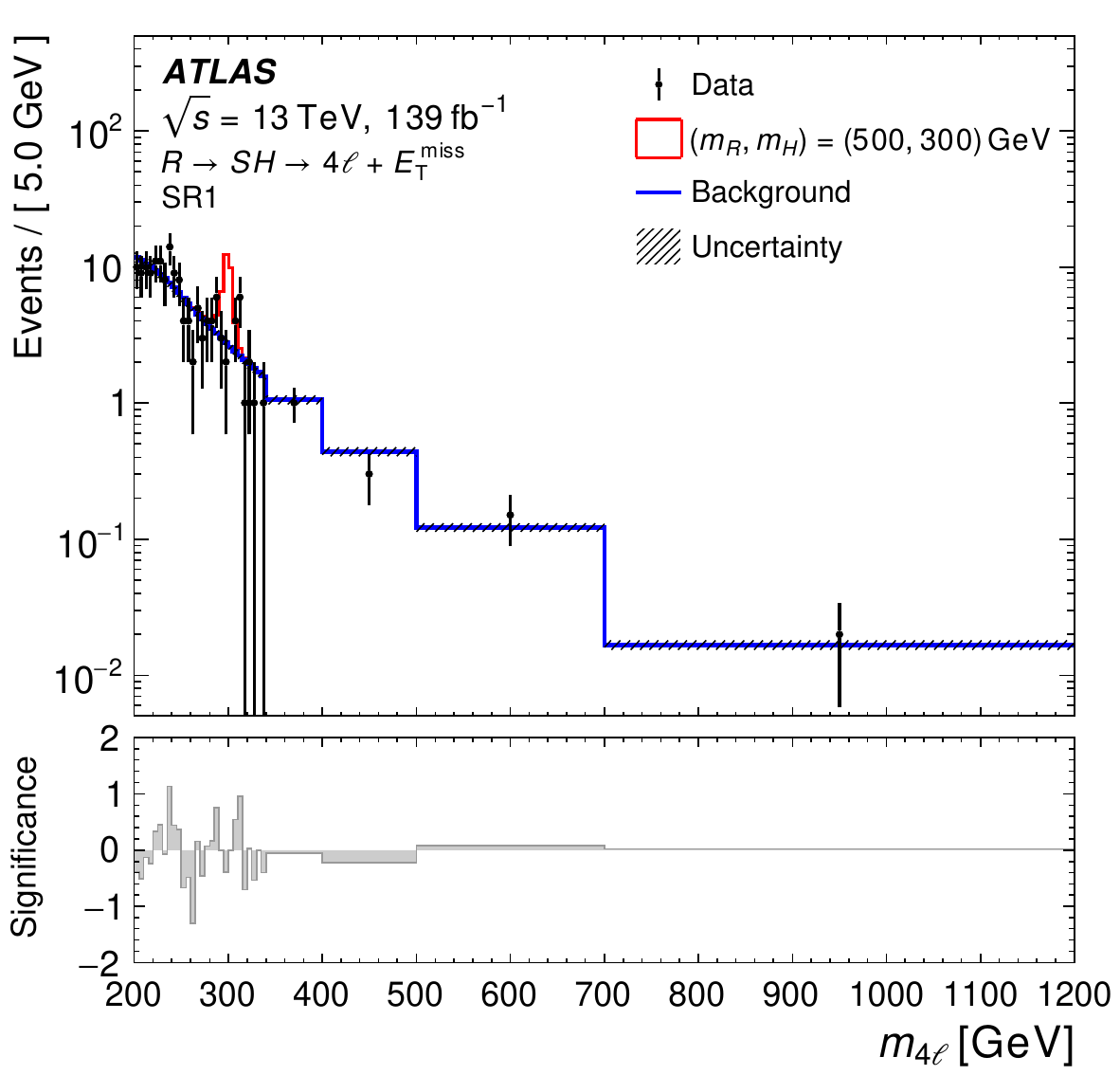}\label{subfig:m4lFit_SR1_RSH}}\;\;
\subfloat[]{\includegraphics[width=0.45\linewidth]{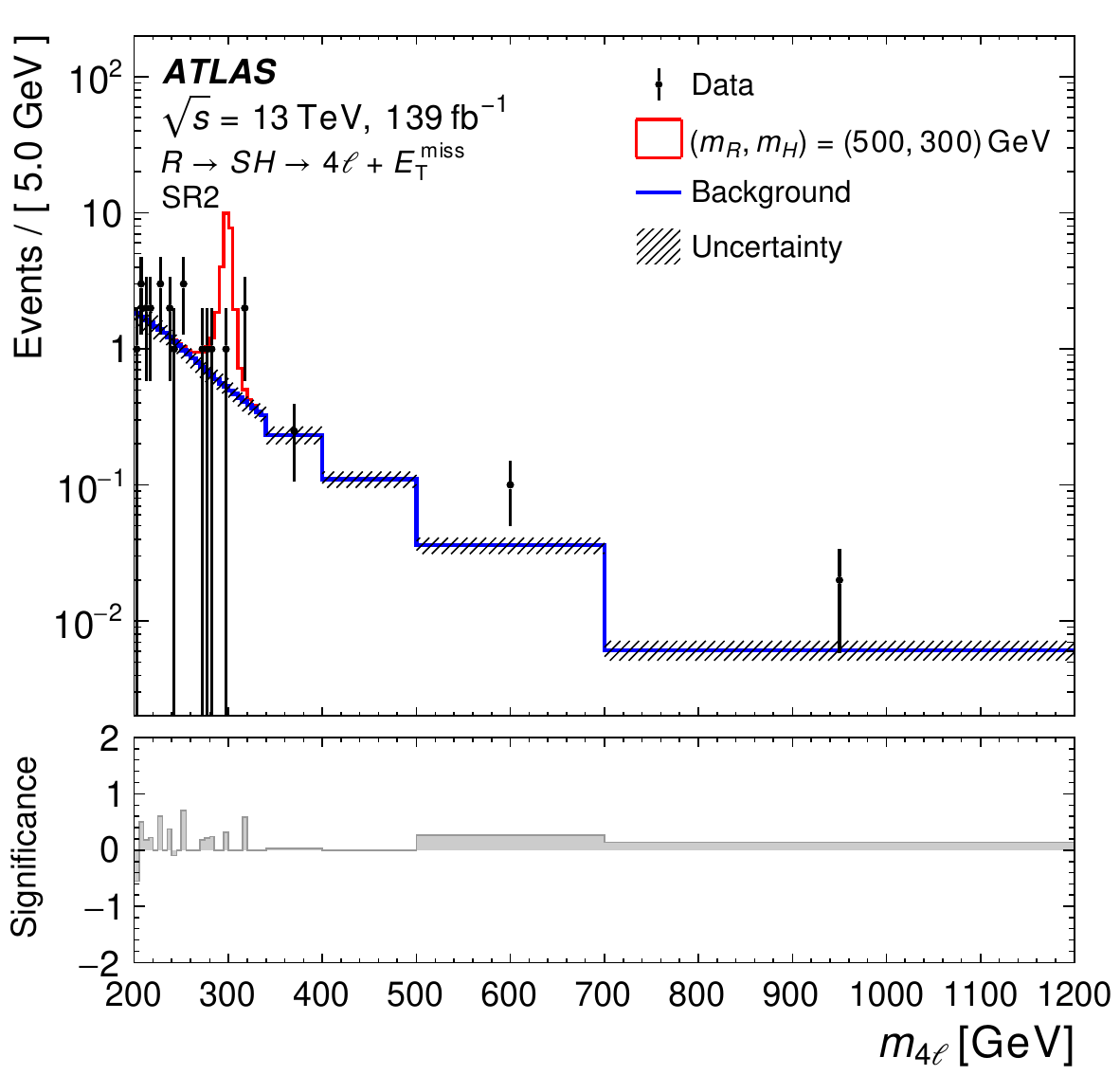}\label{subfig:m4lFit_SR2_RSH}}\\
\subfloat[]{\includegraphics[width=0.45\linewidth]{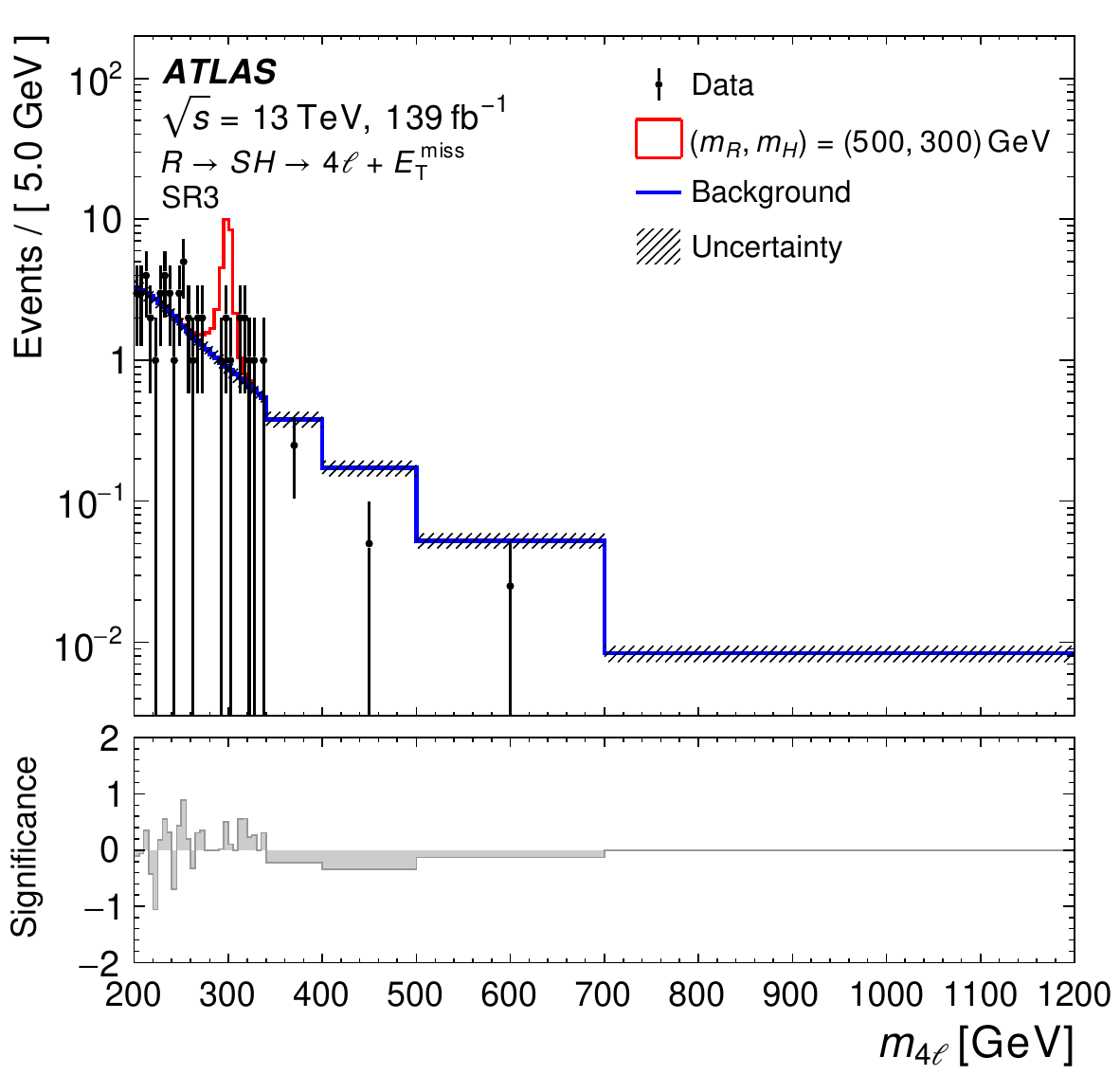}\label{subfig:m4lFit_SR3_RSH}}
\end{center}
\vspace{-0.2cm}
\caption{Observed and expected distributions of the invariant mass of the four-lepton system in the $R\rightarrow SH\rightarrow 4\ell+E^{\textrm{miss}}_{\textrm{T}}$ search for
\protect\subref{subfig:m4lFit_SR1_RSH} SR1, \protect\subref{subfig:m4lFit_SR2_RSH} SR2 and \protect\subref{subfig:m4lFit_SR3_RSH} SR3 under a background-only fit to data. The total background (blue) includes the $q\overline{q}\rightarrow ZZ$, $gg\rightarrow ZZ$, $q\overline{q}\rightarrow ZZ$ (EW), $VVV$,  $t\overline{t}V$, $t\overline{t}$, $Z$+jets and $WZ$ processes. The overlaid distribution of the $(m_{R}, m_{H}) = (500, 300)$~GeV signal (red) is normalised to the observed upper limit on the cross-section (25.0 fb) discussed in Section~\ref{sec:result}. The hatched lines show the systematic uncertainty of the background prediction, while the error bar on the data denote the statistical uncertainty. The lower panel displays the significance of each bin, which is determined by the residual of the data corresponding to the fitted background taking the statistical uncertainty of the data into account. The fit is performed in bins of 5~GeV width, while the plots are re-binned using variable binning for presentation purposes.}\label{m4l_Fit_RSH}
\end{figure}

\begin{figure}[!t]
\begin{center}
\subfloat[]{\includegraphics[width=0.45\linewidth]{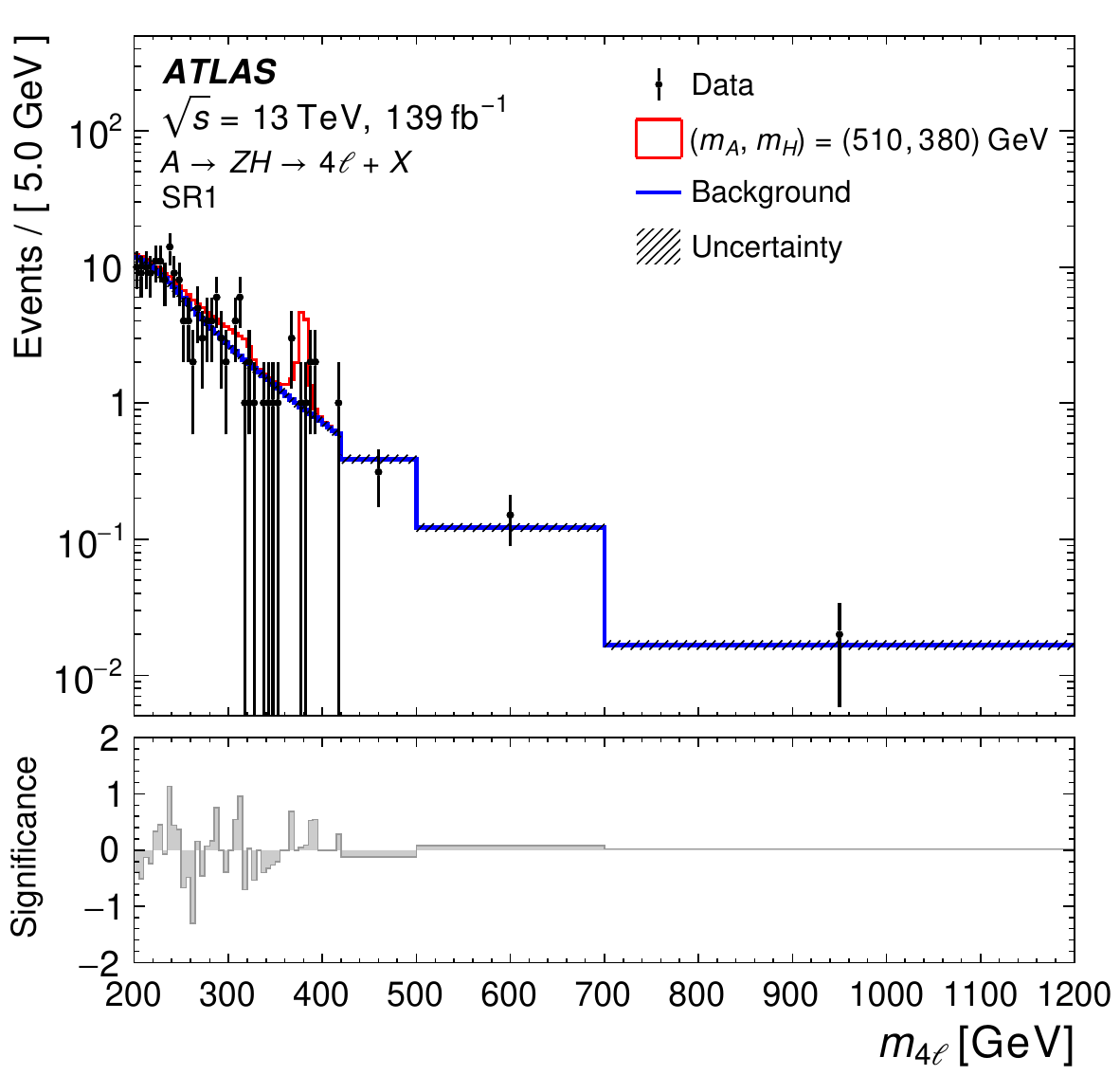}\label{subfig:m4lFit_SR1_AZH}}\;\;
\subfloat[]{\includegraphics[width=0.45\linewidth]{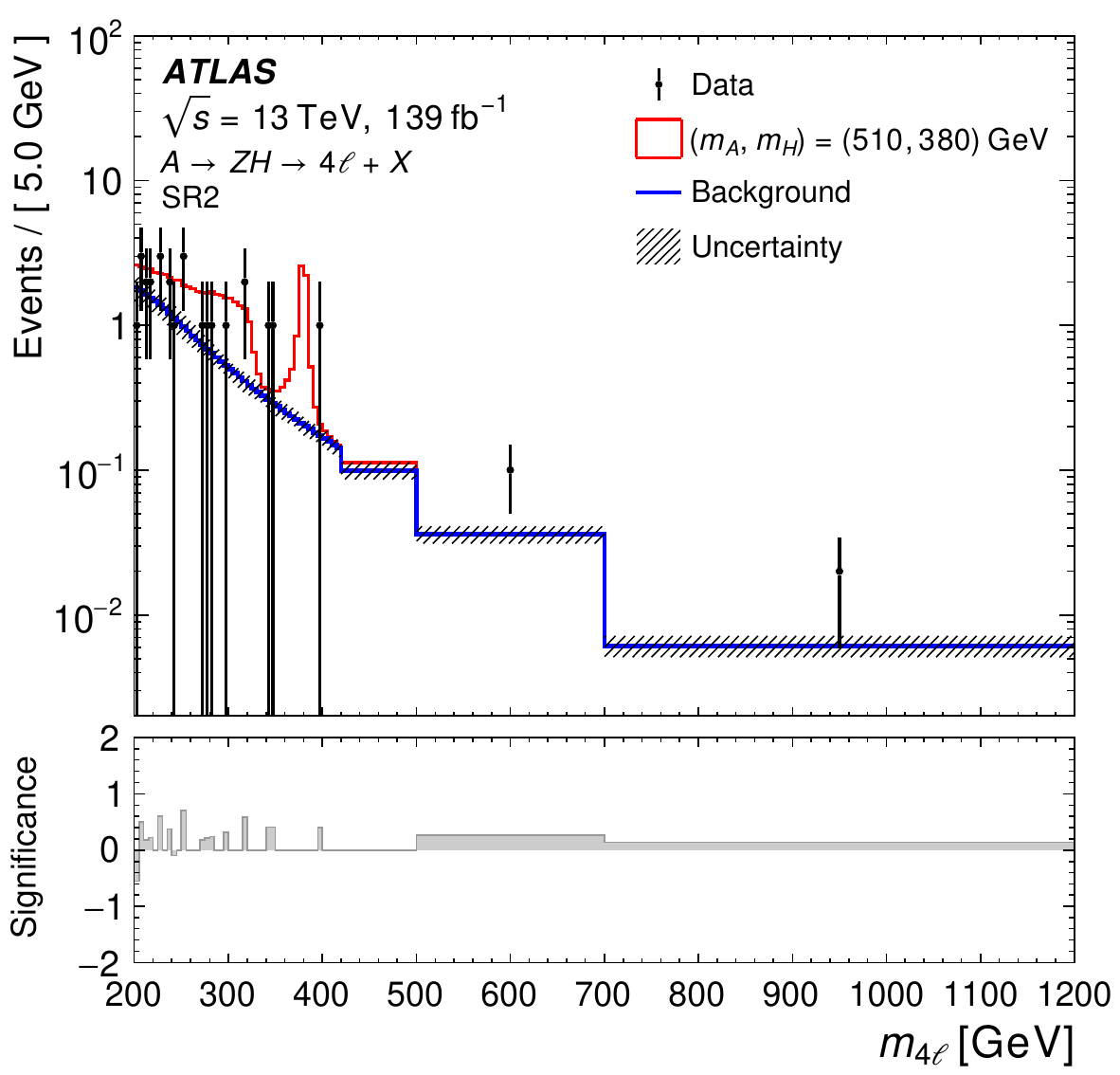}\label{subfig:m4lFit_SR2_AZH}}\\
\subfloat[]{\includegraphics[width=0.45\linewidth]{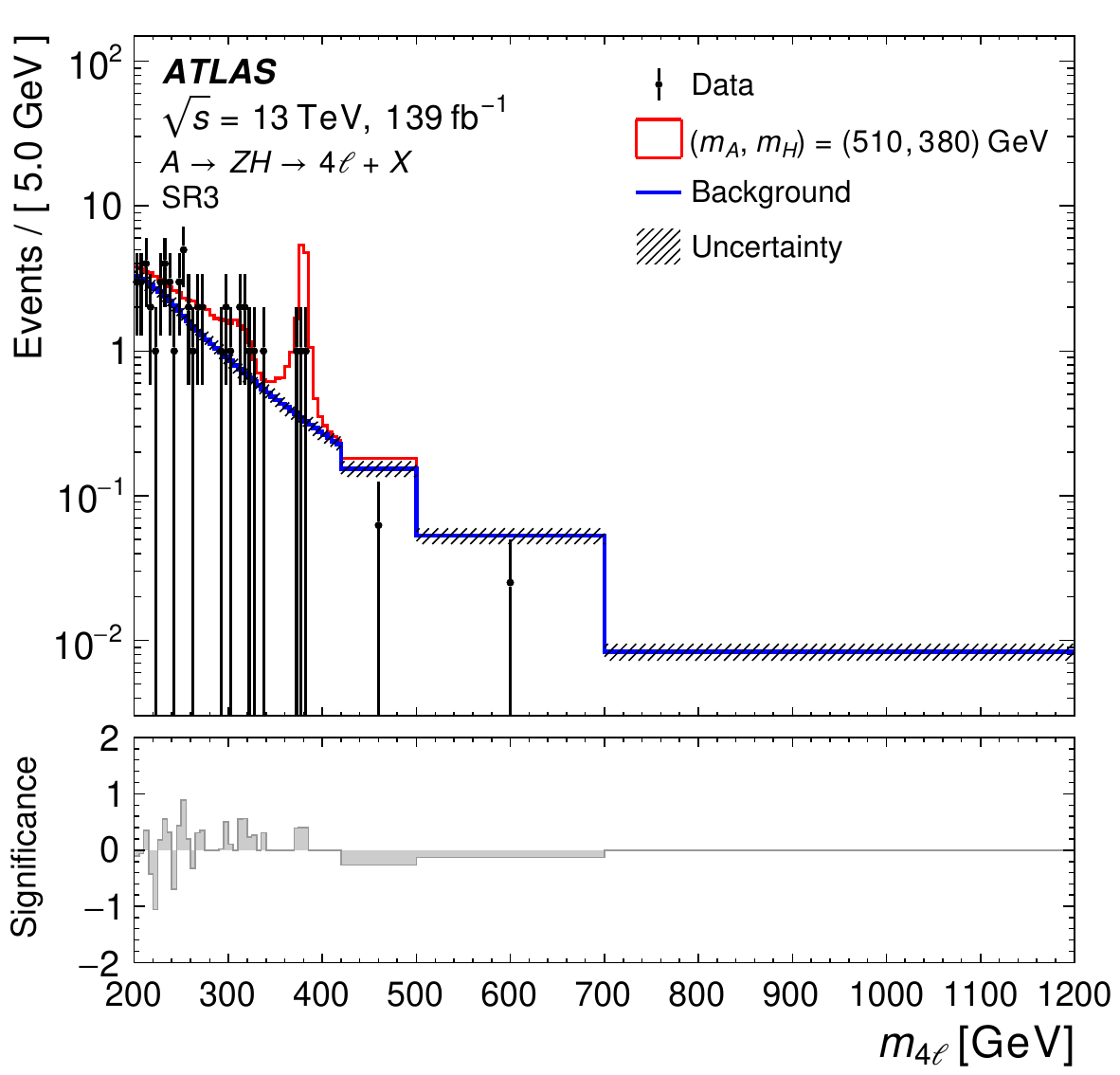}\label{subfig:m4lFit_SR3_AZH}}
\end{center}
\caption{Observed and expected distributions of the invariant mass of the four-lepton system in the $A\rightarrow ZH\rightarrow 4\ell+X$ search for
\protect\subref{subfig:m4lFit_SR1_AZH} SR1, \protect\subref{subfig:m4lFit_SR2_AZH} SR2 and \protect\subref{subfig:m4lFit_SR3_AZH} SR3 under a background-only fit to data. The total background (blue) includes the $q\overline{q}\rightarrow ZZ$, $gg\rightarrow ZZ$, $q\overline{q}\rightarrow ZZ$ (EW), $VVV$,  $t\overline{t}V$, $t\overline{t}$, $Z$+jets and $WZ$ processes. The overlaid distribution of the $(m_{A}, m_{H}) = (510, 380)$~GeV signal (red) is normalised to the observed upper limit on the cross-section (29.9 fb) discussed in Section~\ref{sec:result}. The hatched lines show the systematic uncertainty of the background prediction, while the error bar on the data denote the statistical uncertainty. The lower panel displays the significance of each bin, which is determined by the residual of the data corresponding to the fitted background taking the statistical uncertainty of the data into account. The fit is performed in bins of 5~GeV width, while the plots are re-binned using variable binning for presentation purposes.}\label{fig:m4l_Fit_AZH_1}
\end{figure}

\begin{figure}[!t]
\begin{center}
\subfloat[]{\includegraphics[width=0.45\linewidth]{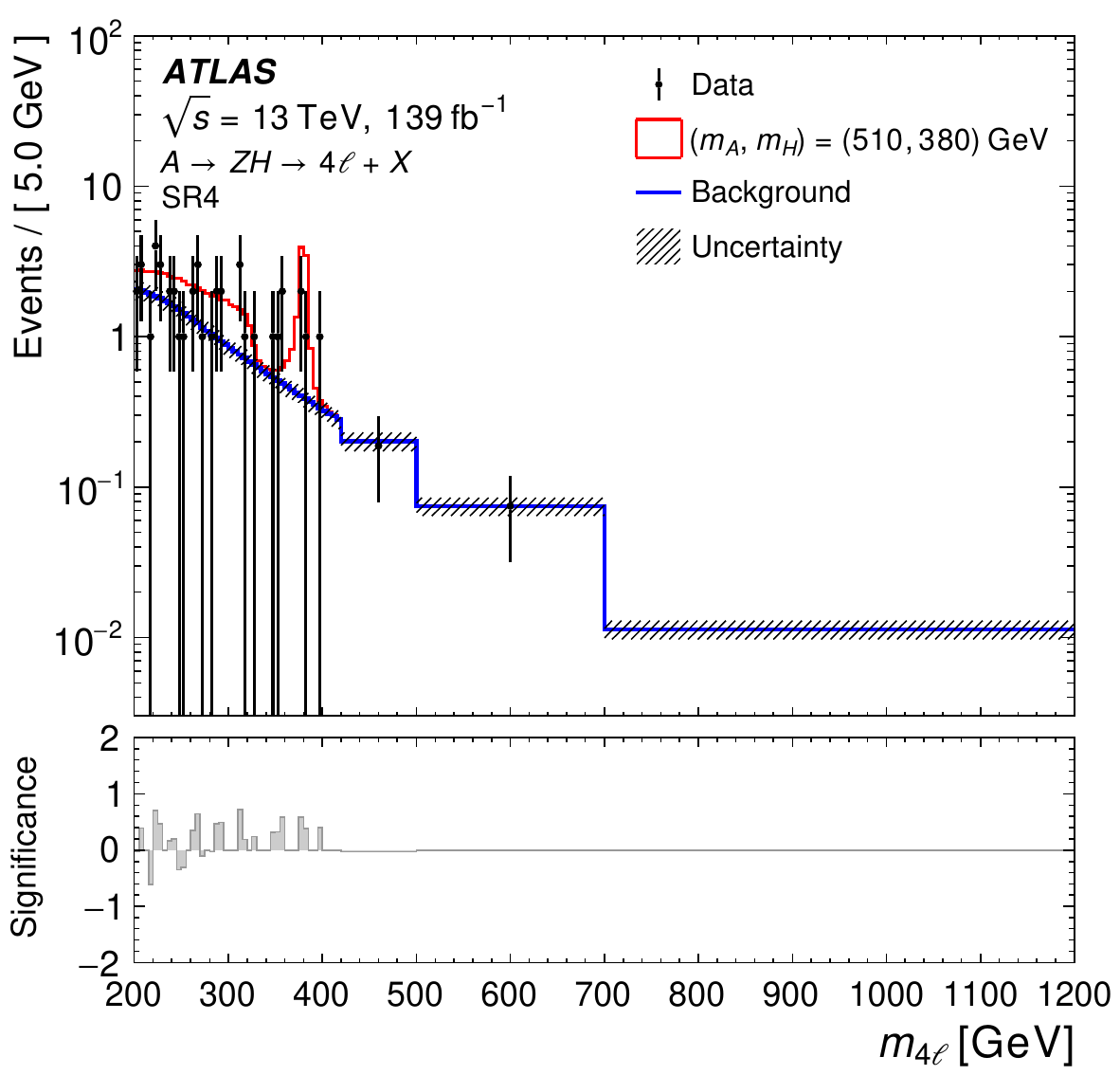}\label{subfig:m4lFit_SR4_AZH}}\;\;
\subfloat[]{\includegraphics[width=0.45\linewidth]{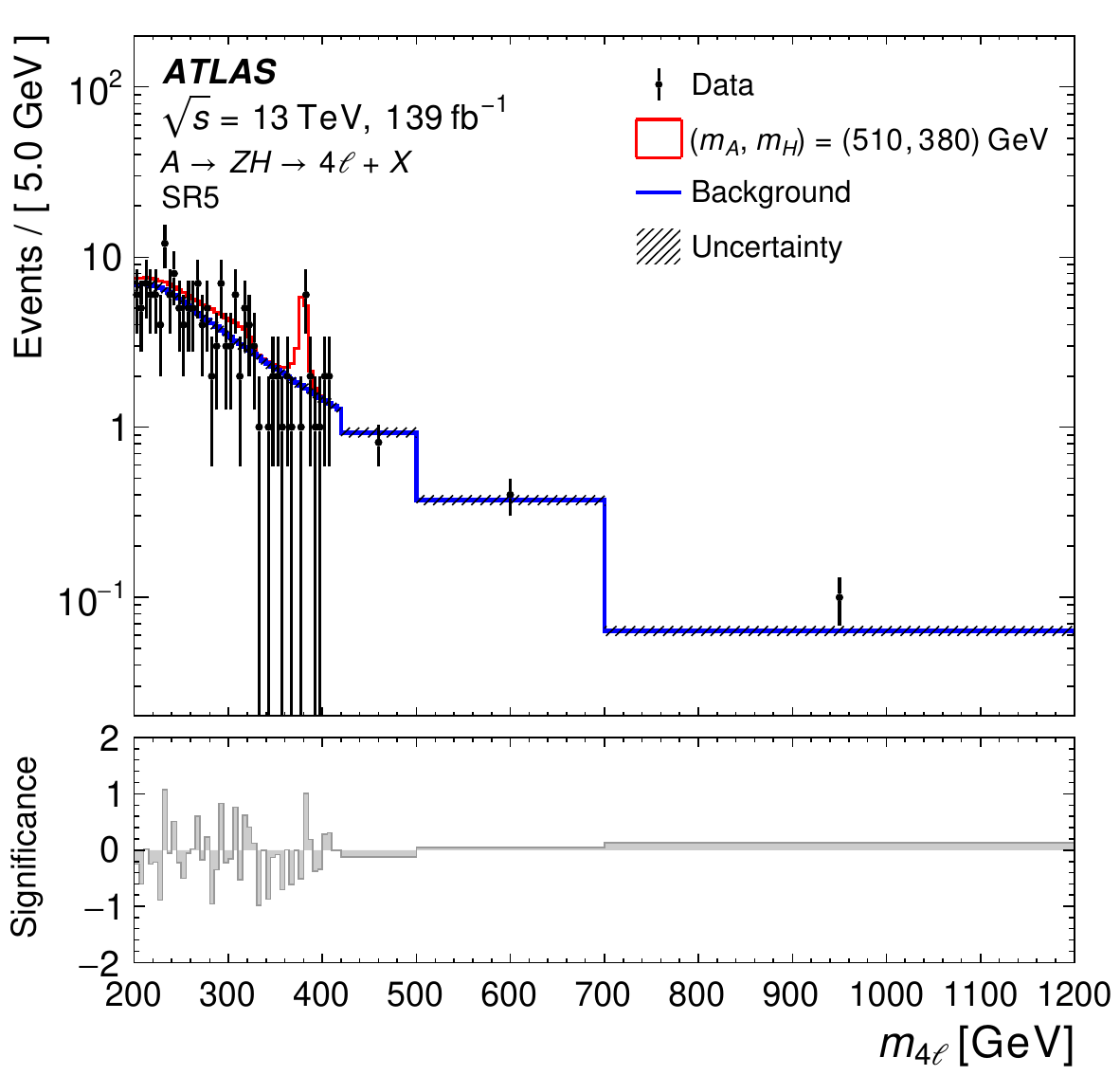}\label{subfig:m4lFit_SR5_AZH}}\\
\subfloat[]{\includegraphics[width=0.45\linewidth]{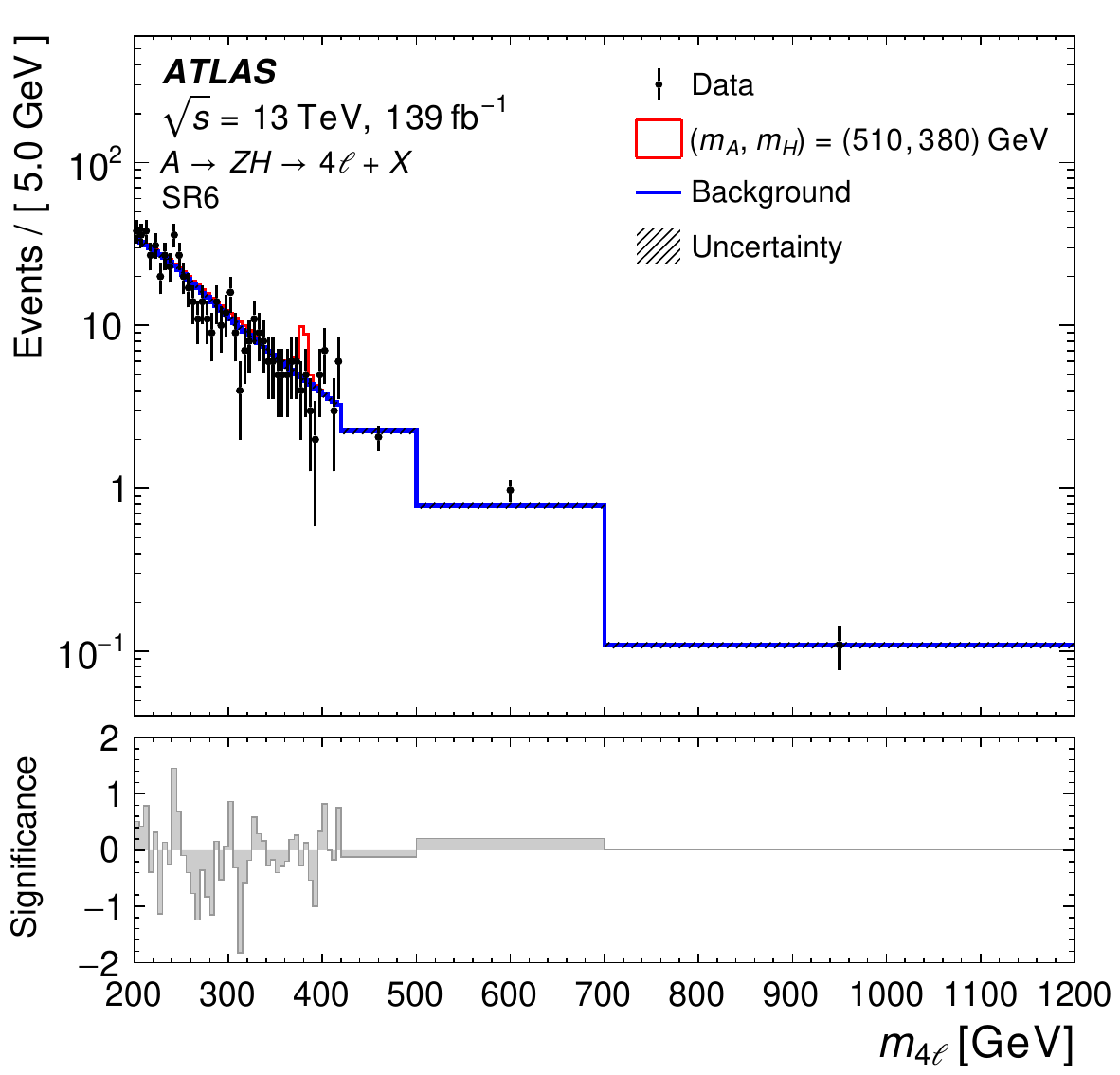}\label{subfig:m4lFit_SR6_AZH}}\;\;
\subfloat[]{\includegraphics[width=0.45\linewidth]{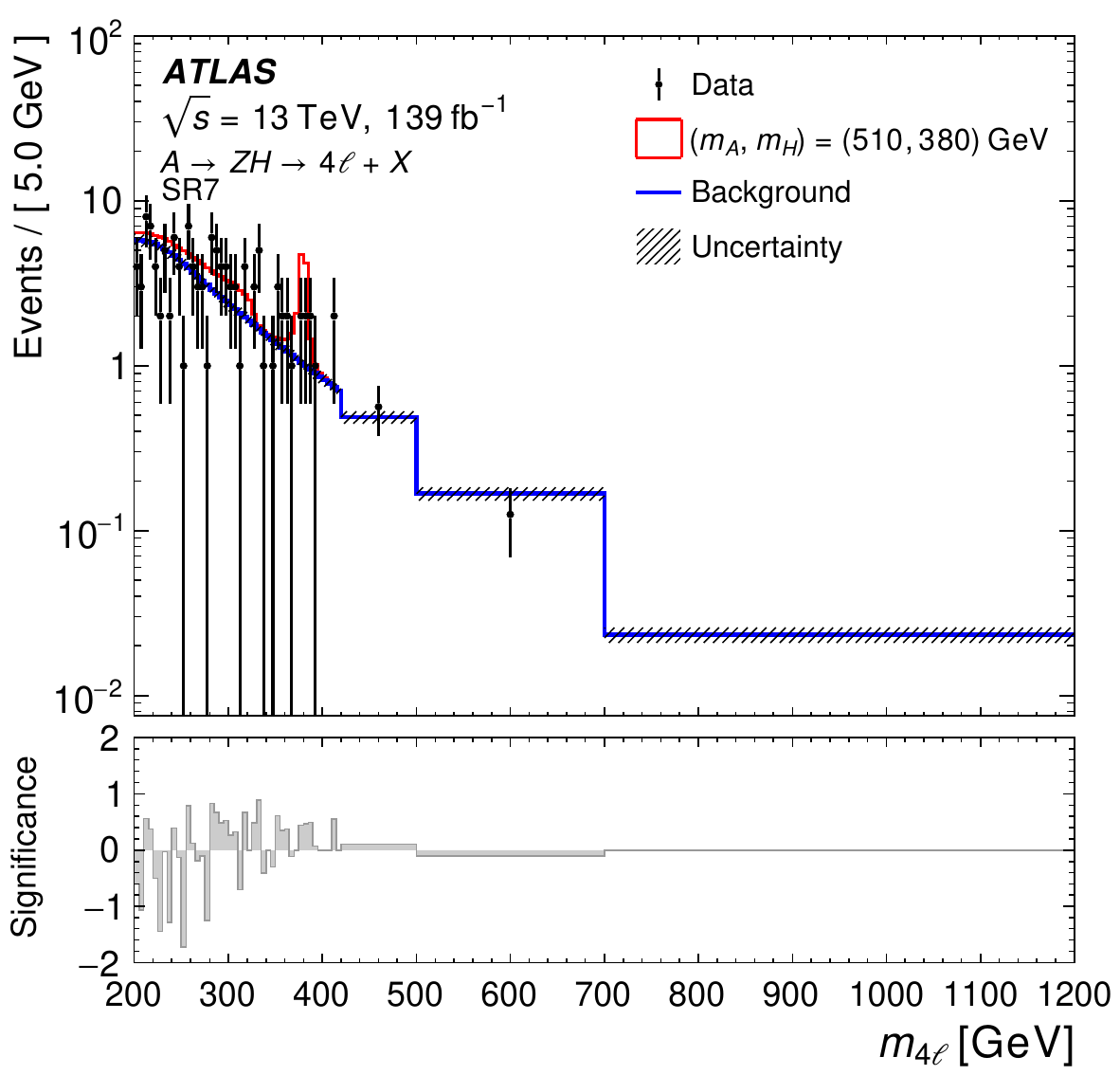}\label{subfig:m4lFit_SR7_AZH}}
\end{center}
\caption{Observed and expected distributions of the invariant mass of the four-lepton system in the $A\rightarrow ZH\rightarrow 4\ell+X$ search for \protect\subref{subfig:m4lFit_SR4_AZH} SR4, \protect\subref{subfig:m4lFit_SR5_AZH} SR5, \protect\subref{subfig:m4lFit_SR6_AZH} SR6 and \protect\subref{subfig:m4lFit_SR7_AZH} SR7 under a background-only fit to data. The total background (blue) includes the $q\overline{q}\rightarrow ZZ$, $gg\rightarrow ZZ$, $q\overline{q}\rightarrow ZZ$ (EW), $VVV$, $t\overline{t}V$, $t\overline{t}$, $Z$+jets and $WZ$ processes. The overlaid distribution of the $(m_{A}, m_{H}) = (510, 380)$~GeV signal (red) is normalised to the observed upper limit on the cross-section (29.9 fb) discussed in Section~\ref{sec:result}. The hatched lines show the systematic uncertainty of the background prediction, while the error bar on the data denote the statistical uncertainty. The lower panel displays the significance of each bin, which is determined by the residual of the data corresponding to the fitted background taking the statistical uncertainty of the data into account. The fit is performed in bins of 5~GeV width, while the plots are re-binned using variable binning for presentation purposes.}\label{m4l_Fit_AZH_2}
\end{figure}

\begin{figure}[!t]
\begin{center}
\subfloat[]{\includegraphics[width=0.49\linewidth]{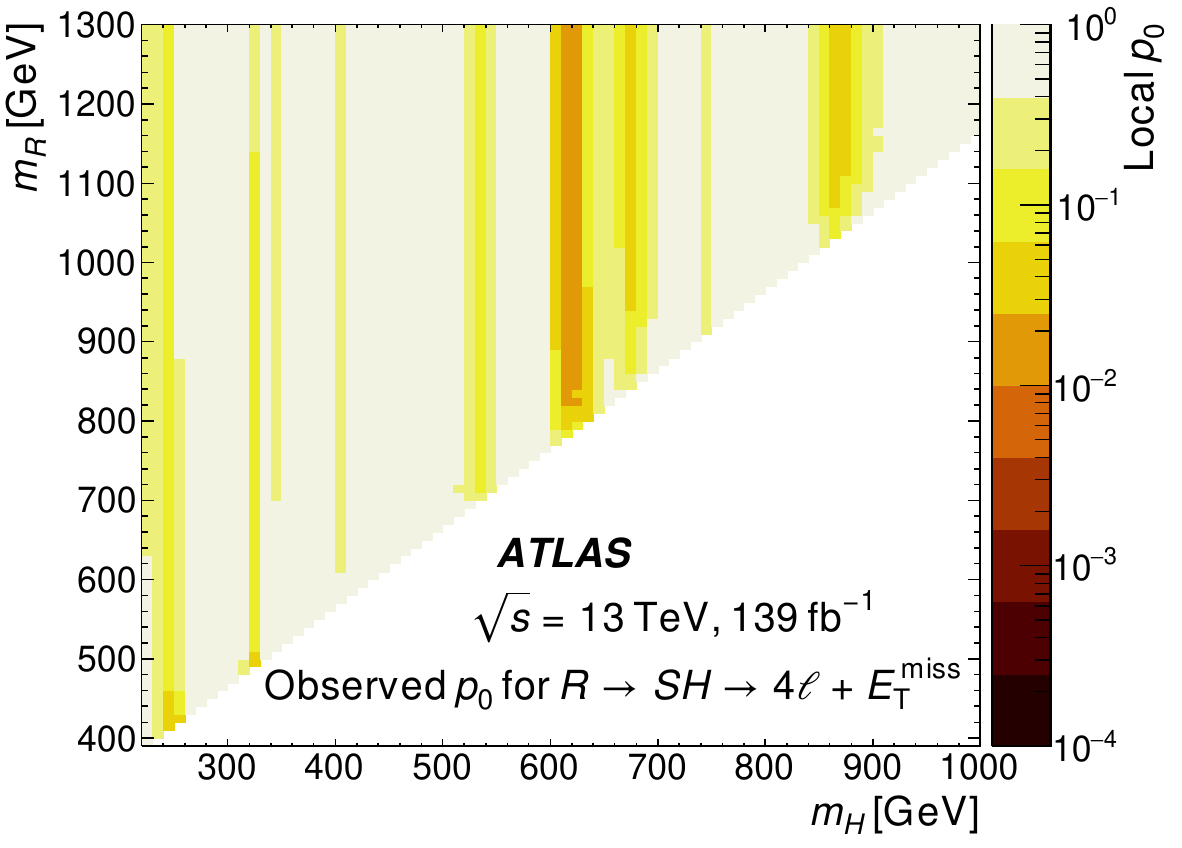}\label{p0_RSH}}\;\;
\subfloat[]{\includegraphics[width=0.49\linewidth]{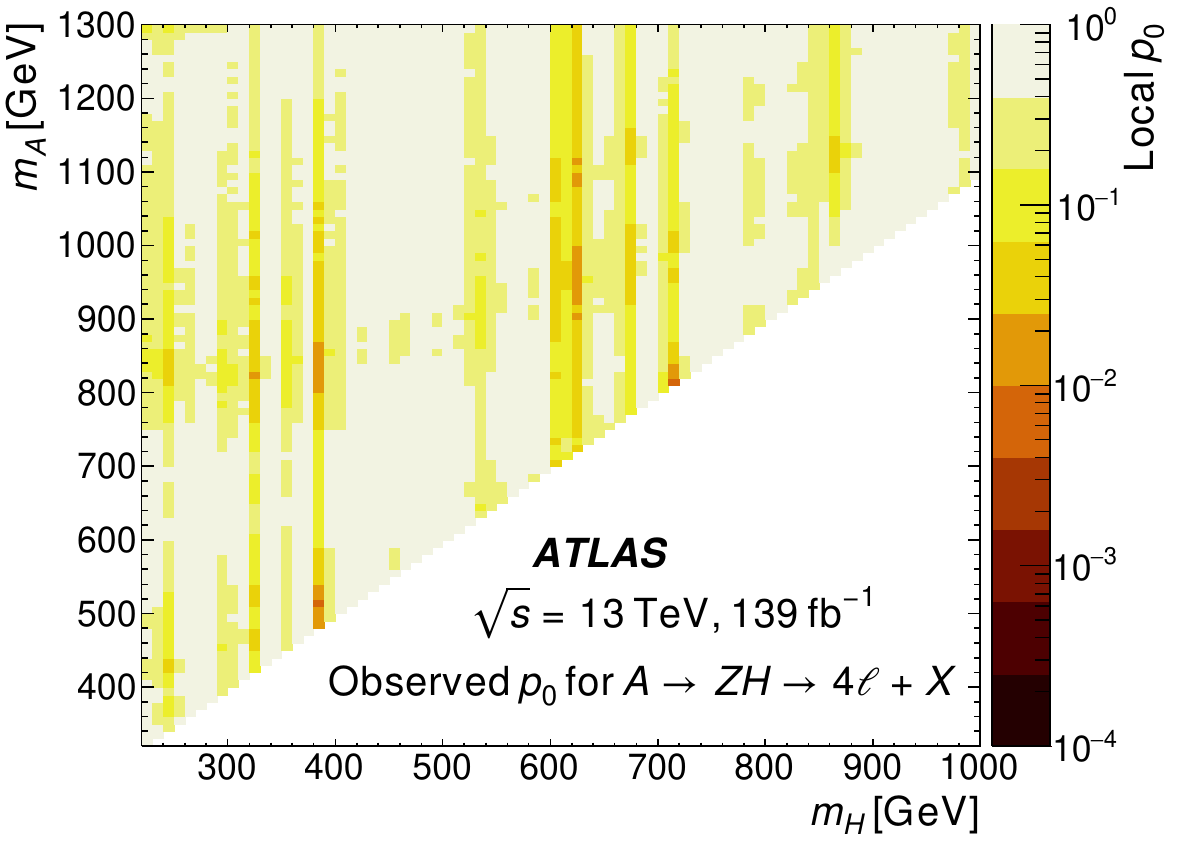}\label{p0_AZH}}
\end{center}
\caption{Local $p_0$-values in the $(m_{H}, m_{R})$ plane for \protect\subref{p0_RSH} the $R\rightarrow SH\rightarrow 4\ell+E^{\textrm{miss}}_{\textrm{T}}$ search with $m_{S} = 160$~GeV, and in the $(m_{H}, m_{A})$ plane for \protect\subref{p0_AZH} the $A\rightarrow ZH\rightarrow 4\ell+X$ search.}
\end{figure}

The four lepton invariant mass distributions for the $R\rightarrow SH\rightarrow 4\ell+E^{\textrm{miss}}_{\textrm{T}}$ search with the $(m_R, m_H) = (500, 300)$~GeV mass point in all three SRs are shown in Figure~\ref{m4l_Fit_RSH}. Figures~\ref{fig:m4l_Fit_AZH_1} and \ref{m4l_Fit_AZH_2} show the $m_{4\ell}$ distribution in the seven SRs defined for the $A\rightarrow ZH\rightarrow 4\ell+X$ search with the $(m_A, m_H) = (510, 380)$~GeV mass point. These plots are shown for a background-only fit to data, and signals are overlaid after being scaled to the observed upper limit on the cross-section. In the presented figures, statistical fluctuations are dominant, particularly in the high mass region of the $m_{4\ell}$. The binning was determined by considering the width and the stability of the upper limit on the cross-section of the $R\rightarrow SH\rightarrow 4\ell+E^{\textrm{miss}}_{\textrm{T}}$ and $A\rightarrow ZH\rightarrow 4\ell+X$ signals. The upper limit calculations were performed in bins with a fixed width of 5~GeV. However, for better visualisation, the plots are presented with variable binning.

No significant deviation from the SM backgrounds is observed in the data. Local excesses around two standard deviations for a few mass points are shown in Figure~\ref{p0_RSH} for the $R\rightarrow SH\rightarrow 4\ell+E^{\textrm{miss}}_{\textrm{T}}$ signal and in Figure \ref{p0_AZH} for the $A\rightarrow ZH\rightarrow 4\ell+X $ signal. The most significant excess comes from the $A\rightarrow ZH\rightarrow 4\ell+X$ signal at the $(m_A, m_H) = (510, 380)$~GeV mass point with a local significance of 2.5 standard deviations. The impact of systematic uncertainties on the analysis is investigated based on their impact on the parameter of interest (best-fit cross-section) for the $R\rightarrow SH\rightarrow 4\ell+E^{\textrm{miss}}_{\textrm{T}}$ and $A\rightarrow ZH\rightarrow 4\ell+X$ signals. The uncertainties that have the most significant impact vary depending on the choice of the $(m_{R}, m_{H})$ or $(m_{A}, m_{H})$ mass points. Table~\ref{Sysimpact} shows the first five leading sources of uncertainties in the best-fit $\sigma$ value for three different mass points for each signal. The effect of the uncertainties is estimated using Asimov data produced with the signal cross-section set to the expected limits for the particular $(m_{R}, m_{H})$ or $(m_{A}, m_{H})$ mass points, assuming narrow width $R$ or $A$ and $H$ bosons. In the $R\rightarrow SH\rightarrow 4\ell+E^{\textrm{miss}}_{\textrm{T}}$ search, systematic uncertainties related to jet characteristics, such as jet flavour composition, played a significant role and ranked first and second at mass points $(m_{R}, m_{H}) = (390, 220)$ GeV and $(m_{R}, m_{H}) = (1300, 1000)$ GeV, respectively. It is noteworthy that the uncertainties associated with jet characteristics remained significant and played a pivotal role in the overall uncertainty assessment for both the $R\rightarrow SH\rightarrow 4\ell+E^{\textrm{miss}}_{\textrm{T}}$ and $A\rightarrow ZH\rightarrow 4\ell+X$ searches.

\begin{table}[t]
\begin{center}
\resizebox{\textwidth}{!}{%
\begin{tabular}{llcllc}
\toprule
\multicolumn{3}{c}{$R\rightarrow SH\rightarrow 4\ell+E^{\textrm{miss}}_{\textrm{T}}$} & \multicolumn{3}{c}{$A\rightarrow ZH\rightarrow 4\ell+X$} \\
\cmidrule(lr){1-3} \cmidrule(l){4-6}
$(m_{R}, m_{H})$ [GeV]    & Uncertainty source  & $\Delta\sigma/\sigma$ [\%]& $(m_{A}, m_{H})$ [GeV]  & Uncertainty source  & $\Delta\sigma/\sigma$ [\%]\\
\cmidrule(lr){1-3} \cmidrule(l){4-6}
&    Jet flavour composition                            &           6.2            &                   & Other backgrounds parameterisation SR7                &    5.0 \\
&    Jet flavour response                               &           4.8            &                   & $q\overline{q}\rightarrow ZZ$ parameterisation SR7    &    3.9 \\
$(390, 220)$   & Jet energy scale                       &           4.2            &   $(320, 220)$    & Jet flavour composition                               &    3.9 \\
&    Pile-up reweighting                                &           4.0            &                   & Luminosity                                            &    3.7 \\
&    CKKW parton showering ($gg\rightarrow ZZ$) SR2     &           3.8            &                   & $gg\rightarrow ZZ$ parameterisation SR6               &    3.6 \\
\cmidrule(lr){1-3} \cmidrule(l){4-6}
&    CKKW parton showering ($gg\rightarrow ZZ$) SR2     &           3.1            &                   & Luminosity                                            &    2.4 \\
&    QSF parton showering ($gg\rightarrow ZZ$) SR2      &           3.0            &                   & Jet flavour composition                               &    2.4 \\
$(500, 300)$        &    $q\overline{q}\rightarrow ZZ$ parameterisation SR2 &           2.0            &   $(510, 380)$    & Jet energy scale                                      &    1.7 \\
&    Pile-up reweighting                                &           1.9            &                   & Jet energy resolution                                 &    1.5 \\
&    $VVV$ parameterisation SR2                         &           1.9            &                   & Signal PDF                                            &    1.4\\
\cmidrule(lr){1-3} \cmidrule(l){4-6}
&    $q\overline{q}\rightarrow ZZ$ parameterisation SR2 &           9.3            &                   & CKKW parton showering ($gg\rightarrow ZZ$) SR2        &    3.3\\
&    Jet flavour composition                            &           7.3            &                   & QSF parton showering ($gg\rightarrow ZZ$) SR2         &    3.3 \\
$(1300, 1000)$      &    Jet flavour response                               &           3.5            &   $(1300, 1000)$  & Other backgrounds parameterisation SR2                &    2.6 \\
&    Pile-up reweighting                                &           2.9            &                   & $q\overline{q}\rightarrow ZZ$ parameterisation SR5    &    2.1 \\
&    CKKW parton showering ($gg\rightarrow ZZ$) SR2     &           2.9            &                   & $VVV$ parameterisation SR2                            &    2.1 \\
\cmidrule(lr){1-6}
\end{tabular}}
\end{center}
\caption{The impact of the most important sources of uncertainty on the best-fit value of the signal cross-section for three mass points for each $R\rightarrow SH\rightarrow 4\ell+E^{\textrm{miss}}_{\textrm{T}}$ and $A\rightarrow ZH\rightarrow 4\ell+X$ signals after the fit. Sum in quadrature of the upward and downward effects ($\Delta\sigma$) is divided by the best fit value of the signal cross-section ($\sigma$) for each signal. Shape uncertainties are uncorrelated and hence affect each region separately.\label{Sysimpact}}
\end{table}

\begin{figure}[t]
\begin{center}
\subfloat[]{\includegraphics[width=0.45\linewidth]{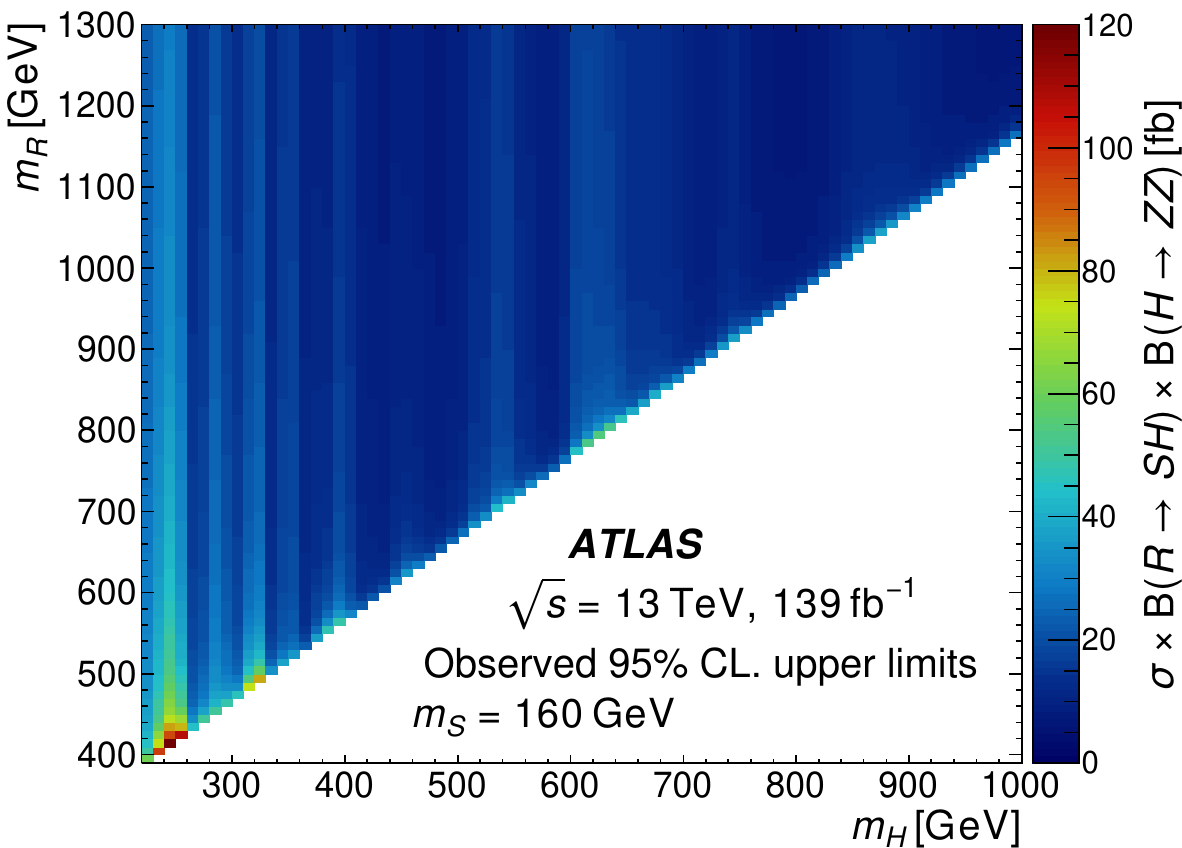}\label{subfig:limit_2D_RSH_obs}}\;\;
\subfloat[]{\includegraphics[width=0.45\linewidth]{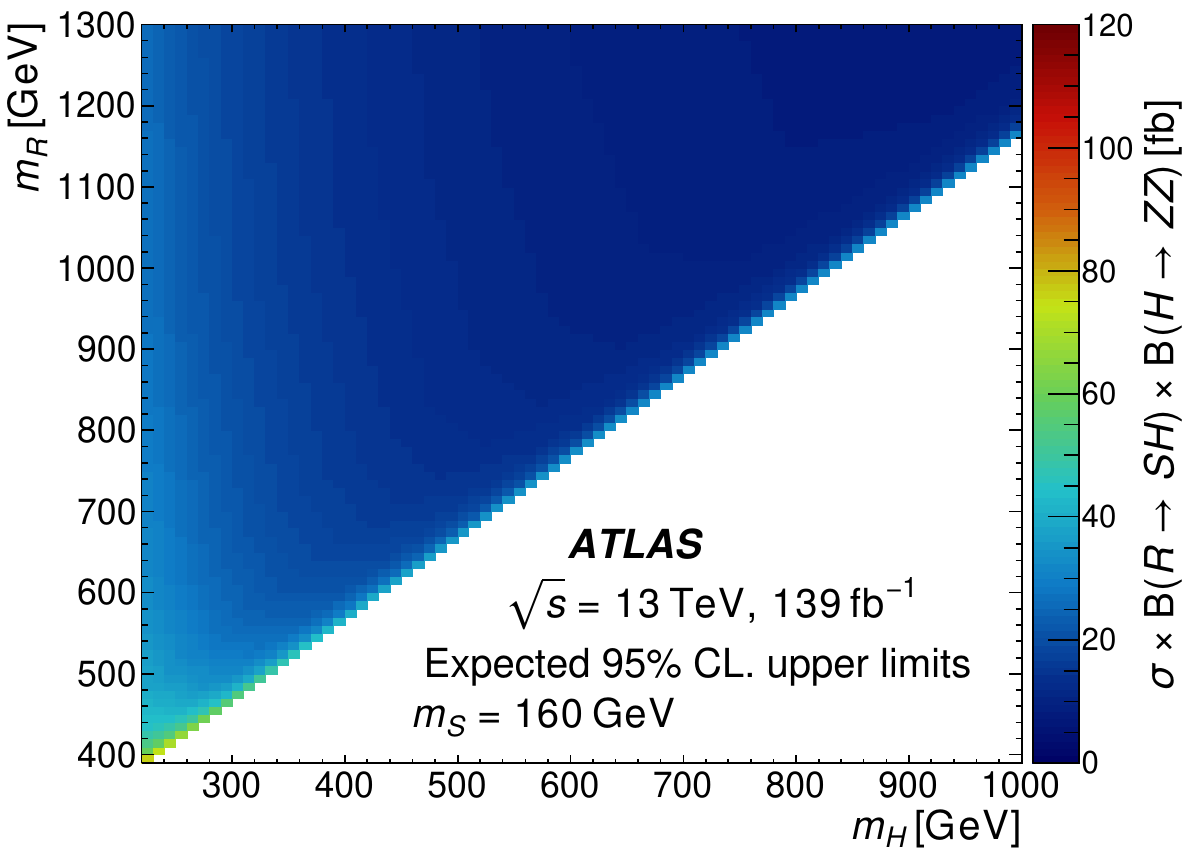}\label{subfig:limit_2D_RSH_exp}}\\
\subfloat[]{\includegraphics[width=0.45\linewidth]{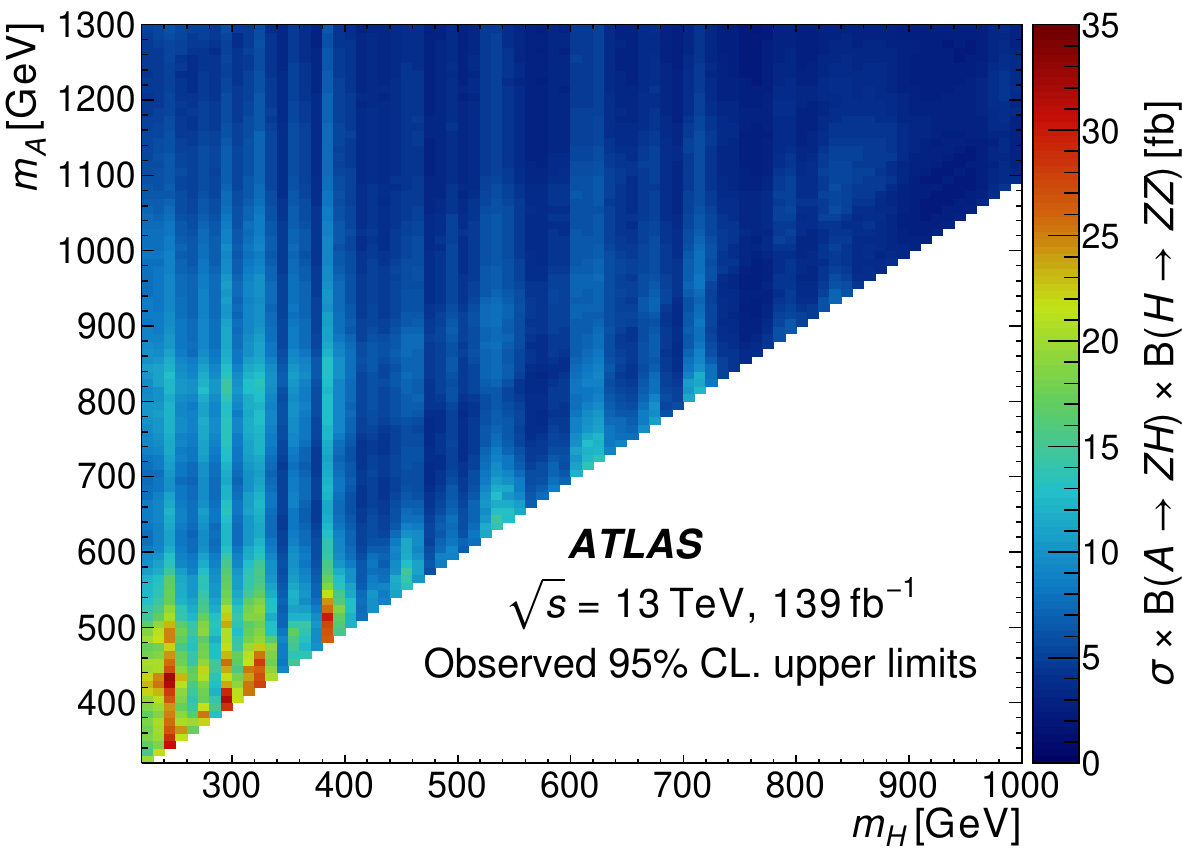}\label{subfig:limit_2D_AZH_obs}}\;\;
\subfloat[]{\includegraphics[width=0.45\linewidth]{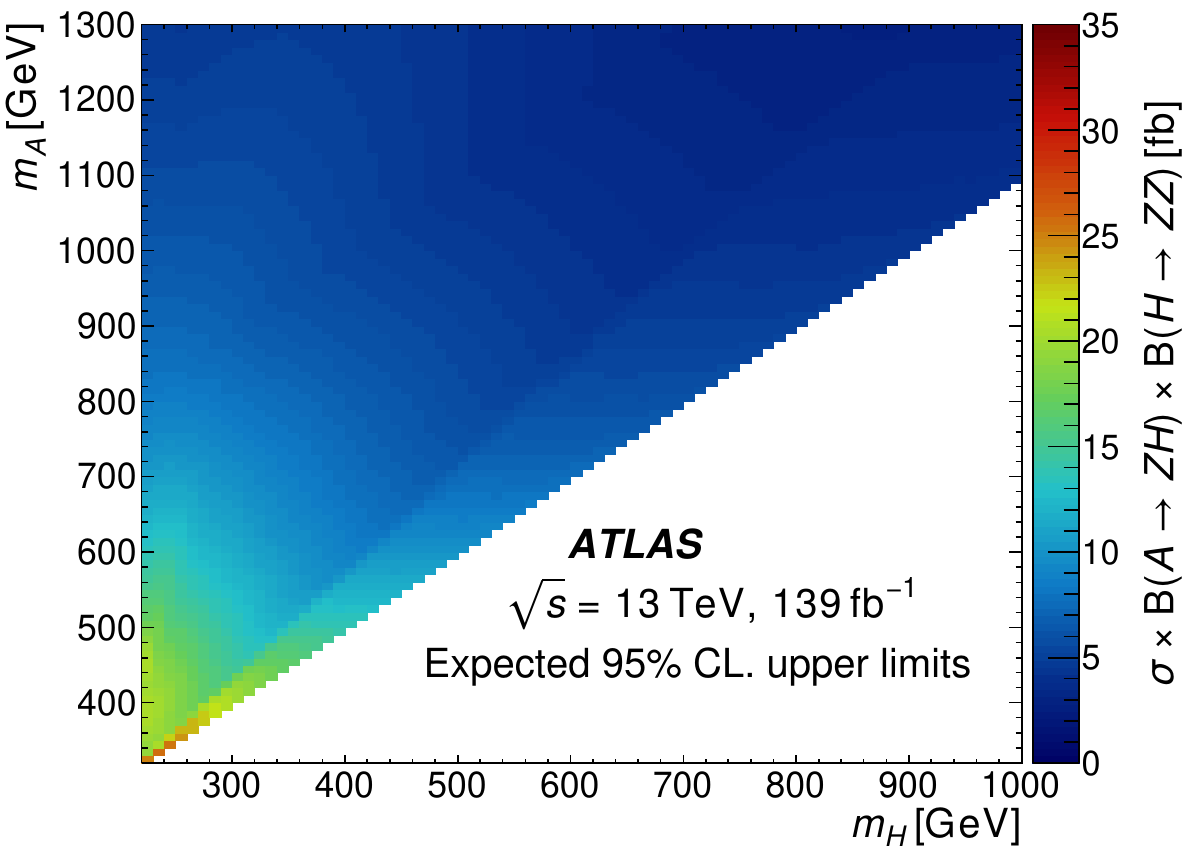}\label{subfig:limit_2D_AZH_exp}}
\end{center}
\caption{The observed (left) and expected (right) upper limits at 95\% confidence level on \protect\subref{subfig:limit_2D_RSH_obs}--\protect\subref{subfig:limit_2D_RSH_exp} $\sigma(gg\rightarrow R)\times\mathcal{B}(R\rightarrow SH)\times\mathcal{B}(H\rightarrow ZZ)$ across the $(m_H,\,m_R$) plane with $m_S=160$~GeV for the $R\rightarrow SH\rightarrow 4\ell+E^{\textrm{miss}}_{\textrm{T}}$ search, and on \protect\subref{subfig:limit_2D_AZH_obs}--\protect\subref{subfig:limit_2D_AZH_exp} $\sigma(gg\rightarrow A)\times\mathcal{B}(A\rightarrow ZH)\times\mathcal{B}(H\rightarrow ZZ)$ across the $(m_H,\,m_A$) plane for the $A\rightarrow ZH\rightarrow 4\ell+X$ search.}
\end{figure}

As no significant excess was observed in comparison to the background predictions, the results were translated into upper limits on the production cross-section of the $R\rightarrow SH\rightarrow 4\ell+E^{\textrm{miss}}_{\textrm{T}}$ and $A\rightarrow ZH\rightarrow 4\ell+X$ signals times the branching fractions. The branching fractions considered are $\mathcal{B}(R\rightarrow SH)$ and $\mathcal{B}(H\rightarrow ZZ)$ for the $R\rightarrow SH\rightarrow 4\ell+E^{\textrm{miss}}_{\textrm{T}} $ process. For the $A\rightarrow ZH\rightarrow 4\ell+X$ process, the branching fractions are $\mathcal{B}(A\rightarrow ZH)$ and $\mathcal{B}(H\rightarrow ZZ)$. The CL$_s$ approach in the asymptotic approximation~\cite{Cowan:2010js, Read:2002hq} was used to calculate upper limits at 95\% confidence level (CL) across the explored phase space. The upper limit, either on $\sigma(gg\rightarrow R)\times\mathcal{B}(R\rightarrow SH)\times\mathcal{B}(H\rightarrow ZZ)$ or on $\sigma(gg\rightarrow A)\times\mathcal{B}(A\rightarrow ZH)\times\mathcal{B}(H\rightarrow ZZ)$, for a specific mass hypothesis, is obtained by fixing the $H$ mass parameter to a constant value and optimising the probability function for nuisance parameters. The observed and expected upper limits of NWA $R$ and $H$ bosons for the $R\rightarrow SH\rightarrow 4\ell+E^{\textrm{miss}}_{\textrm{T}}$ search are shown in Figures~\ref{subfig:limit_2D_RSH_obs} and \ref{subfig:limit_2D_RSH_exp} in the $(m_{H}, m_{R})$ plane where the $z$-axis displays the upper limits on $\sigma(gg\rightarrow R)\times\mathcal{B}(R\rightarrow SH)\times\mathcal{B}(H\rightarrow ZZ)$ in fb. The observed upper limits for the $R\rightarrow SH\rightarrow 4\ell+E^{\textrm{miss}}_{\textrm{T}}$ search range from 6.8 fb for $(m_R, m_H) = (1300, 980)$~GeV to 119.2 fb for $(m_R, m_H) = (410, 240)$~GeV. In contrast, the expected upper limits vary from 7.7 fb to 70.3 fb for the same mass points. Similarly, Figures~\ref{subfig:limit_2D_AZH_obs} and \ref{subfig:limit_2D_AZH_exp} show the observed and expected upper limits for the $A\rightarrow ZH\rightarrow 4\ell+X$ search in the $(m_H, m_A)$ plane, in which the $z$-axis represents the upper limits on $\sigma(gg\rightarrow A)\times\mathcal{B}(A\rightarrow ZH)\times\mathcal{B}(H\rightarrow ZZ)$. For the $A\rightarrow ZH\rightarrow 4\ell+X$ search, the observed upper limits range from 2.1 fb for $(m_A, m_H) = (1300, 760)$~GeV to 32.3 fb for $(m_A, m_H) = (430, 240)$~GeV, whereas the corresponding expected upper limits range from 2.9 fb to 18.8 fb. Pseudo-experiments were used to calculate upper limits for various mass points, but the results did not significantly deviate from the asymptotic upper limits.

The analysis focuses on studying heavy resonances with NWA $A$ or $R$ and $H$ bosons produced via gluon--fusion for the $A\rightarrow ZH\rightarrow 4\ell+X$ and $R\rightarrow SH\rightarrow 4\ell+E^{\textrm{miss}}_{\textrm{T}}$ processes. However, the effect of the LWA in a similar scenario was also checked. For this purpose, two low and high mass points, $(m_A, m_H) = (320, 220)$~GeV and $(m_A, m_H) = (1190, 600)$~GeV, for the $A\rightarrow ZH\rightarrow 4\ell+X$ signal were generated using similar configurations as for the NWA signal. Upper limits were computed for signal assumptions in which the $A$ and $H$ bosons have a significant natural width relative to the true mass. They were found to increase as the natural width of the $A$ and $H$ bosons increases. For example, in the case of an $A\rightarrow ZH\rightarrow 4\ell+X$ signal that produces $A$ and $H$ bosons with natural widths of 15\% and 5\%, respectively, of their true mass, the upper limit worsened by a factor of 1.4 for the low mass point and 1.7 for the high mass point compared to the NWA upper limit. In the $(\Gamma_A/m_A, \Gamma_H/m_H) = (30\%, 10\%)$ case, the upper limit is worsened by a factor of 1.7 for the low mass point and by a factor of 1.9 for the high mass point when compared to the NWA scenario. Table~\ref{tab:limiLW} summarises width assumptions, mass points and upper limits for LWA and NWA signals.
\begin{table*}%
\begin{center}
\resizebox{\textwidth}{!}{%
\begin{tabular}{lcccc}
\toprule
\multirow{2}{*}{Width assumptions} & \multirow{2}{*}{Mass points [GeV]} & \multicolumn{2}{c}{Upper limits in the $\sigma(gg\rightarrow A)$ [fb]} & \multirow{2}{*}{\begin{tabular}[c]{@{}l@{}}Ratio w.r.t\\ Narrow width\end{tabular}} \\ \cmidrule(lr){3-4}
&                                                             & Observed                             & Expected                            &                           \\ \midrule
\multirow{2}{*}{Narrow width} & $(m_A, m_H) = (320, 220)$     & 19.6 & 25.1 & 1.0  \\\cmidrule(l){2-2}
& $(m_A, m_H) = (1190, 600)$    & \pho4.8 & \pho3.5 & 1.0  \\
\midrule
\multirow{2}{*}{$(\Gamma_A /m_A , \Gamma_H /m_H )=  (15\%, 5\%)$}  & $(m_A, m_H) = (320, 220)$  & 31.5 & 36.2 & 1.4 \\ \cmidrule(l){2-2}
& $(m_A, m_H) = (1190, 600)$ & \pho8.3 & \pho6.0 & 1.7 \\
\midrule
\multirow{2}{*}{$(\Gamma_A /m_A , \Gamma_H /m_H )=  (30\%, 10\%)$} & $(m_A, m_H) = (320, 220)$  & 38.9 & 42.5 & 1.7 \\\cmidrule(l){2-2}
& $(m_A, m_H) = (1190, 600)$ & \pho8.9 & \pho6.6 & 1.9 \\
\bottomrule
\end{tabular}}
\end{center}
\caption{Observed and expected upper limits at 95\% confidence level (CL) on the $\sigma(gg\rightarrow A)\times\mathcal{B}(A\rightarrow ZH)\times\mathcal{B}(H\rightarrow ZZ)$ for different large width approximation signals for comparison with the narrow width approximation. The fraction in the rightmost column corresponds to the ratio between the expected upper limits with large- or narrow width signals. The $\Gamma_A$ and $\Gamma_H$ denote the widths of the $A$ and $H$ bosons, respectively.}\label{tab:limiLW}
\end{table*}


%
%
\FloatBarrier
\section{Conclusion}
\label{sec:conclusion}

A search for new heavy resonances is performed in a final state with four leptons and missing transverse energy or jets. The search uses proton-proton collision data at a centre-of-mass energy of 13 TeV collected by the ATLAS detector from 2015 to 2018 at the Large Hadron Collider corresponding to a total integrated luminosity of 139 fb$^{-1}$. The search focuses on two signal processes: $R\rightarrow SH\rightarrow 4\ell+E^{\textrm{miss}}_{\textrm{T}}$ and $A\rightarrow ZH\rightarrow 4\ell+X$. For the $R\rightarrow SH\rightarrow 4\ell+E^{\textrm{miss}}_{\textrm{T}}$ search, the $R$ boson decays to $S$ and $H$ bosons, where $S\rightarrow \,\textrm{invisible}$ and $H\rightarrow ZZ\rightarrow 4\ell$. In the $A\rightarrow ZH\rightarrow 4\ell+X$ search, the $A$ boson decays to $Z$ and $H$ bosons, with two possible decay modes considered for the associated production of $Z$ and $H$ bosons: $A\rightarrow Z(\rightarrow jj/\ell^+\ell^-/\textrm{invisible})H(\rightarrow 4\ell)$ and $A\rightarrow Z(\rightarrow 2\ell)H(\rightarrow 2\ell+jj/\textrm{invisible})$. For both the $A\rightarrow ZH\rightarrow 4\ell+X$ and $R\rightarrow SH\rightarrow 4\ell+E^{\textrm{miss}}_{\textrm{T}}$ processes, the heavy boson $H$ mass is between 220~GeV and 1000~GeV. The mass range considered for the hypothetical resonance is in the range of 320--1300~GeV for the $A$ boson and 390--1300~GeV for the $R$ boson, while the $S$ boson mass is fixed at 160~GeV, without any significant impact on the results.

No significant deviation above the SM backgrounds is observed. The results are translated into upper limits on $\sigma(gg\rightarrow R)\times\mathcal{B}(R\rightarrow SH)\times\mathcal{B}(H\rightarrow ZZ)$ and $\sigma(gg\rightarrow A)\times\mathcal{B}(A\rightarrow ZH)\times\mathcal{B}(H\rightarrow ZZ)$ at 95\% confidence level. For the $R\rightarrow SH\rightarrow 4\ell+E^{\textrm{miss}}_{\textrm{T}}$ search, the observed (expected) upper limits range from 6.8--119.2 (7.6--75.8) fb for $(m_R, m_H) = (390, 220)$~GeV to $(m_R, m_H) = (1300, 1000)$~GeV. For the $A\rightarrow ZH\rightarrow 4\ell+X$ search, the observed (expected) upper limits range from 2.1--32.3 (2.7--25.8) fb for $(m_{A}, m_{H}) = (320, 220)$~GeV to $(m_{A}, m_{H}) = (1300, 1000)$ GeV. This complements previous $A\rightarrow ZH$ searches in various channels (as discussed in Section~\ref{sec:intro}) and contributes to a more comprehensive understanding of the behaviour of $A$ and $H$ bosons. Furthermore, the $R\rightarrow SH\rightarrow 4\ell+E^{\textrm{miss}}_{\textrm{T}}$ search extends the gluon-fusion production of the $H\rightarrow ZZ\rightarrow 4\ell$ inclusive analysis, incorporating the possibility of missing transverse energy in the four-lepton final states and extending the phase-space of the search.

%


%
\section*{Acknowledgements}
%

%
%

%
%

We thank CERN for the very successful operation of the LHC and its injectors, as well as the support staff at
CERN and at our institutions worldwide without whom ATLAS could not be operated efficiently.

The crucial computing support from all WLCG partners is acknowledged gratefully, in particular from CERN, the ATLAS Tier-1 facilities at TRIUMF/SFU (Canada), NDGF (Denmark, Norway, Sweden), CC-IN2P3 (France), KIT/GridKA (Germany), INFN-CNAF (Italy), NL-T1 (Netherlands), PIC (Spain), RAL (UK) and BNL (USA), the Tier-2 facilities worldwide and large non-WLCG resource providers. Major contributors of computing resources are listed in Ref.~\cite{ATL-SOFT-PUB-2023-001}.

We gratefully acknowledge the support of ANPCyT, Argentina; YerPhI, Armenia; ARC, Australia; BMWFW and FWF, Austria; ANAS, Azerbaijan; CNPq and FAPESP, Brazil; NSERC, NRC and CFI, Canada; CERN; ANID, Chile; CAS, MOST and NSFC, China; Minciencias, Colombia; MEYS CR, Czech Republic; DNRF and DNSRC, Denmark; IN2P3-CNRS and CEA-DRF/IRFU, France; SRNSFG, Georgia; BMBF, HGF and MPG, Germany; GSRI, Greece; RGC and Hong Kong SAR, China; ISF and Benoziyo Center, Israel; INFN, Italy; MEXT and JSPS, Japan; CNRST, Morocco; NWO, Netherlands; RCN, Norway; MEiN, Poland; FCT, Portugal; MNE/IFA, Romania; MESTD, Serbia; MSSR, Slovakia; ARIS and MVZI, Slovenia; DSI/NRF, South Africa; MICINN, Spain; SRC and Wallenberg Foundation, Sweden; SERI, SNSF and Cantons of Bern and Geneva, Switzerland; NSTC, Taipei; TENMAK, T\"urkiye; STFC/UKRI, United Kingdom; DOE and NSF, United States of America.

Individual groups and members have received support from BCKDF, CANARIE, CRC and DRAC, Canada; CERN-CZ, PRIMUS 21/SCI/017 and UNCE SCI/013, Czech Republic; COST, ERC, ERDF, Horizon 2020, ICSC-NextGenerationEU and Marie Sk{\l}odowska-Curie Actions, European Union; Investissements d'Avenir Labex, Investissements d'Avenir Idex and ANR, France; DFG and AvH Foundation, Germany; Herakleitos, Thales and Aristeia programmes co-financed by EU-ESF and the Greek NSRF, Greece; BSF-NSF and MINERVA, Israel; Norwegian Financial Mechanism 2014-2021, Norway; NCN and NAWA, Poland; La Caixa Banking Foundation, CERCA Programme Generalitat de Catalunya and PROMETEO and GenT Programmes Generalitat Valenciana, Spain; G\"{o}ran Gustafssons Stiftelse, Sweden; The Royal Society and Leverhulme Trust, United Kingdom.

In addition, individual members wish to acknowledge support from CERN: European Organization for Nuclear Research (CERN PJAS); Chile: Agencia Nacional de Investigaci\'on y Desarrollo (FONDECYT 1190886, FONDECYT 1210400, FONDECYT 1230812, FONDECYT 1230987); China: National Natural Science Foundation of China (NSFC - 12175119, NSFC 12275265, NSFC-12075060); Czech Republic: PRIMUS Research Programme (PRIMUS/21/SCI/017); EU: H2020 European Research Council (ERC - 101002463); European Union: European Research Council (ERC - 948254, ERC 101089007), Horizon 2020 Framework Programme (MUCCA - CHIST-ERA-19-XAI-00), European Union, Future Artificial Intelligence Research (FAIR-NextGenerationEU PE00000013), Italian Center for High Performance Computing, Big Data and Quantum Computing (ICSC, NextGenerationEU); France: Agence Nationale de la Recherche (ANR-20-CE31-0013, ANR-21-CE31-0013, ANR-21-CE31-0022, ANR-22-EDIR-0002), Investissements d'Avenir Labex (ANR-11-LABX-0012); Germany: Baden-Württemberg Stiftung (BW Stiftung-Postdoc Eliteprogramme), Deutsche Forschungsgemeinschaft (DFG - 469666862, DFG - CR 312/5-1); Italy: Istituto Nazionale di Fisica Nucleare (ICSC, NextGenerationEU), Ministero dell'Università e della Ricerca (PRIN - 20223N7F8K - PNRR M4.C2.1.1); Japan: Japan Society for the Promotion of Science (JSPS KAKENHI JP21H05085, JSPS KAKENHI JP22H01227, JSPS KAKENHI JP22H04944, JSPS KAKENHI JP22KK0227); Netherlands: Netherlands Organisation for Scientific Research (NWO Veni 2020 - VI.Veni.202.179); Norway: Research Council of Norway (RCN-314472); Poland: Polish National Agency for Academic Exchange (PPN/PPO/2020/1/00002/U/00001), Polish National Science Centre (NCN 2021/42/E/ST2/00350, NCN OPUS nr 2022/47/B/ST2/03059, NCN UMO-2019/34/E/ST2/00393, UMO-2020/37/B/ST2/01043, UMO-2021/40/C/ST2/00187, UMO-2022/47/O/ST2/00148); Slovenia: Slovenian Research Agency (ARIS grant J1-3010); Spain: BBVA Foundation (LEO22-1-603), Generalitat Valenciana (Artemisa, FEDER, IDIFEDER/2018/048), La Caixa Banking Foundation (LCF/BQ/PI20/11760025), Ministry of Science and Innovation (MCIN \& NextGenEU PCI2022-135018-2, MICIN \& FEDER PID2021-125273NB, RYC2019-028510-I, RYC2020-030254-I, RYC2021-031273-I, RYC2022-038164-I), PROMETEO and GenT Programmes Generalitat Valenciana (CIDEGENT/2019/023, CIDEGENT/2019/027); Sweden: Swedish Research Council (VR 2018-00482, VR 2022-03845, VR 2022-04683, VR grant 2021-03651), Knut and Alice Wallenberg Foundation (KAW 2017.0100, KAW 2018.0157, KAW 2018.0458, KAW 2019.0447); Switzerland: Swiss National Science Foundation (SNSF - PCEFP2\_194658); United Kingdom: Leverhulme Trust (Leverhulme Trust RPG-2020-004); United States of America: U.S. Department of Energy (ECA DE-AC02-76SF00515), Neubauer Family Foundation.

%
%


%
%
%
%
%
%
%

%
%
%
%
%
%
%
%
%
%
\clearpage
\printbibliography
\clearpage
 
\begin{flushleft}
\hypersetup{urlcolor=black}
{\Large The ATLAS Collaboration}

\bigskip

\AtlasOrcid[0000-0002-6665-4934]{G.~Aad}$^\textrm{\scriptsize 102}$,
\AtlasOrcid[0000-0001-7616-1554]{E.~Aakvaag}$^\textrm{\scriptsize 16}$,
\AtlasOrcid[0000-0002-5888-2734]{B.~Abbott}$^\textrm{\scriptsize 120}$,
\AtlasOrcid[0000-0002-1002-1652]{K.~Abeling}$^\textrm{\scriptsize 55}$,
\AtlasOrcid[0000-0001-5763-2760]{N.J.~Abicht}$^\textrm{\scriptsize 49}$,
\AtlasOrcid[0000-0002-8496-9294]{S.H.~Abidi}$^\textrm{\scriptsize 29}$,
\AtlasOrcid[0000-0002-9987-2292]{A.~Aboulhorma}$^\textrm{\scriptsize 35e}$,
\AtlasOrcid[0000-0001-5329-6640]{H.~Abramowicz}$^\textrm{\scriptsize 151}$,
\AtlasOrcid[0000-0002-1599-2896]{H.~Abreu}$^\textrm{\scriptsize 150}$,
\AtlasOrcid[0000-0003-0403-3697]{Y.~Abulaiti}$^\textrm{\scriptsize 117}$
% ...
\bigskip
\\

% ...
$^{16}$Department for Physics and Technology, University of Bergen, Bergen; Norway.\\
% ...
$^{29}$Physics Department, Brookhaven National Laboratory, Upton NY; United States of America.\\
% ...
$^{35}$$^{(a)}$Facult\'e des Sciences Ain Chock, Universit\'e Hassan II de Casablanca;$^{(b)}$Facult\'{e} des Sciences, Universit\'{e} Ibn-Tofail, K\'{e}nitra;$^{(c)}$Facult\'e des Sciences Semlalia, Universit\'e Cadi Ayyad, LPHEA-Marrakech;$^{(d)}$LPMR, Facult\'e des Sciences, Universit\'e Mohamed Premier, Oujda;$^{(e)}$Facult\'e des sciences, Universit\'e Mohammed V, Rabat;$^{(f)}$Institute of Applied Physics, Mohammed VI Polytechnic University, Ben Guerir; Morocco.\\
% ...
$^{49}$Fakult\"{a}t Physik , Technische Universit{\"a}t Dortmund, Dortmund; Germany.\\
% ...
$^{55}$II. Physikalisches Institut, Georg-August-Universit\"{a}t G\"ottingen, G\"ottingen; Germany.\\
% ...
$^{102}$CPPM, Aix-Marseille Universit\'e, CNRS/IN2P3, Marseille; France.\\
% ...
$^{117}$Department of Physics, New York University, New York NY; United States of America.\\
% ...
$^{120}$Homer L. Dodge Department of Physics and Astronomy, University of Oklahoma, Norman OK; United States of America.\\
% ...
$^{150}$Department of Physics, Technion, Israel Institute of Technology, Haifa; Israel.\\
$^{151}$Raymond and Beverly Sackler School of Physics and Astronomy, Tel Aviv University, Tel Aviv; Israel.\\
% ...
$^{a}$ Also Affiliated with an institute covered by a cooperation agreement with CERN.\\
$^{b}$ Also at An-Najah National University, Nablus; Palestine.\\
$^{c}$ Also at Borough of Manhattan Community College, City University of New York, New York NY; United States of America.\\
$^{d}$ Also at Center for High Energy Physics, Peking University; China.\\
$^{e}$ Also at Center for Interdisciplinary Research and Innovation (CIRI-AUTH), Thessaloniki; Greece.\\
$^{f}$ Also at Centro Studi e Ricerche Enrico Fermi; Italy.\\
% ...

\end{flushleft}


%

%
%
\clearpage

\end{document}